\documentclass{jfm}
\usepackage{graphicx}
\usepackage{epstopdf, epsfig}
\usepackage{color}

\shorttitle{DBM for non-equilibrium flows}
\shortauthor{Y. Zhang, A. Xu, G. Zhang, Z. Chen and P. Wang}

\title{Discrete Boltzmann method for non-equilibrium flows: based on Shakhov model}

\author{Yudong Zhang\aff{1,2},
  Aiguo Xu\aff{1,3}
   \corresp{\email{Xu\_Aiguo@iapcm.ac.cn}},
  Guangcai Zhang\aff{1},
  Zhihua Chen\aff{2},
 \and Pei Wang\aff{1}
 }

\affiliation{\aff{1}National Key Laboratory of Computational Physics, Institute of Applied Physics and
Computational Mathematics, P. O. Box 8009-26, Beijing 100088, China
\aff{2}Key Laboratory of Transient Physics, Nanjing University of Science and Technology, Nanjing 210094, China
\aff{3}Center for Applied Physics and Technology, MOE Key Center for High Energy Density
Physics Simulations, College of Engineering, Peking University, Beijing 100871, China
}
\begin{document}

\maketitle

\begin{abstract}
A general framework for constructing discrete Boltzmann model for non-equilibrium flows based on the Shakhov model is presented. The Hermite polynomial expansion and a set of discrete velocity with isotropy are adopted to solve the kinetic moments of discrete equilibrium distribution function. Such a model possesses both an adjustable specific heat ratio and Prandtl number, and can be applied to a wide range of flow regimes including continuous, slip, and transition flows. To recover results for actual situations, the nondimensionalization process is demonstrated. To verify and validate the new model, several typical non-equilibrium flows including the Couette flow, Fourier flow, unsteady boundary heating problem, cavity flow, and Kelvin-Helmholtz instability are simulated. Comparisons are made between the results of discrete Boltzmann model and those of previous models including analytic solution in slip flow, Lattice ES-BGK, and DSMC based on both BGK and hard-sphere models. The results show that the new model can accurately capture the velocity slip and temperature jump near the wall, and show excellent performance in predicting the non-equilibrium flow even in transition flow regime. In addition, the measurement of non-equilibrium effects is further developed and the non-equilibrium strength $D^*_n$ in the $n$-th order moment space is defined. The non-equilibrium characteristics and the advantage of using $D^*_n$ in Kelvin-Helmholtz instability are discussed. It concludes that the non-equilibrium strength $D^*_n$ is more appropriate to describe the interfaces than the individual components of $\pmb{\Delta}^*_n$. Besides, the $D^*_3$ and $D^*_{3,1}$ can provide higher resolution interfaces in the simulation of Kelvin-Helmholtz instability.

\end{abstract}


\section{Introduction}
The hypersonic rarefied gas flows with strong nonequilibrium characteristics are often encountered when spacecraft reentry into the atmosphere because of the low density of air at high altitude\citep{Tsien1946Superaerodynamics,Li2015Rarefied}. Similar non-equilibrium flows can also happen in industrial processes in which low density gases are involved, such as the design of vacuum pump. Besides, with the development of microtechnologies, such as Micro-Electro-Mechanical System (MEMS), microfluidic devices\citep{Ho1998MICRO,Stone2004Engineering}, and modern material processing technologies including laser fabrication processing and plasma etching\citep{Gottscho1992Microscopic,Sugioka2012Femtosecond}, gas flows in micro-scale geometries have attracted greatly attention. Those kinds of flows also show significant non-equilibrium effects due to the small characteristic length. In addition, the thermodynamic states around the shock wavefront and detonation wavefront are also far from equilibrium because the width wavefront is only about a few molecular average free path. All of the non-equilibrium flows mentioned above challenge the validity of the traditional hydrodynamic equations.


Generally, rarefied effects of gas flows are characterized by the Knudsen number which is defined as $Kn=\lambda / L$, where $\lambda$ is the mean free path of molecule and $L$ is a representative length scale in the flow. From this sense, \emph{the Knudsen number is generally used to describe the continuity or discreteness/sparsity of the flow system}. According to the values of $Kn$, flow regimes can be divided into continuous flow ($Kn<0.001$), slip flow ($0.001<Kn<0.1$), transitional flow ($0.1<Kn<10$), and free molecular flow ($Kn>10$) \citep{Tsien1946Superaerodynamics,ChingShen2005Rarefied}.
Since based on the continuum hypothesis, Navier-Stokes(NS) equations are only applicable to the continuous flow regime. Although the NS equations can also be extended to slip flow regime, special treatments on the boundary including velocity slip and temperature jump are needed \citep{Manela2010Gas}. Even so, the velocity and temperature profiles within the Knudsen layer near the wall can not be described by the NS equations \citep{Sonebook2007}. In order to effectively capture the flow behaviors within the Knudsen layer, the geometry-dependent viscosity and extended Navier-Stokes constitution are introduced \citep[see][]{Guo2007An}.
Besides, in transitional regime, the constitutive equations, i.e., the Newton's viscosity law and Fourier heat conduction law in NS equations, are not applicable anymore, thus higher order hydrodynamic models such as Burnett equations \citep{Burnett1935The} and Grad's 13-moment equations \citep{Grad1949} were proposed.

In practical terms, the Knudsen numbers in micro-scale flows often cover so wide a range that the flow characteristics can not be well described by only one set of hydrodynamic equations. As the fundamental equation of nonequilibrium statistical physics, the Boltzmann equation possesses the potential to describe the whole range of Knudsen number. From this sense, the Boltzmann equation works for flows with multi-scale structures.   Unfortunately, the original Boltzmann equation is too complicated to be solved directly because of its high dimensional integral collision term. In general, there are two kinds of numerical methods, the probabilistic method and deterministic method, to simulate systems described by Boltzmann equation. As a well known example of probabilistic method, direct simulation Monte Carlo (DSMC) method has been well developed and widely used in rarefied gas dynamics for high speed flows \citep{Bird2003,ChingShen2005Rarefied}. However, it is too expensive in terms of computational cost for low speed flows which are more common in micro-scale flows. Although some improved algorithms, such as the information preservation (IP) method \citep{Fan2001Statistical} and low-variance deviational simulation Monte Carlo (LVDSMC) \citep{Homolle2007A}, have been presented, the contradiction between the noise ratio and computational efficiency has not been well resolved yet.

For deterministic methods, to reduce the computational cost great efforts have been made and various simplified schemes have been developed. Examples are referred to the fast spectral method (FSM) \citep{Wu2013Deterministic}, unified gas-kinetic scheme (UGKS) \citep{UGKS1,UGKS2}, the discrete unified gas-kinetic scheme (DUGKS) \citep{DUGKS1,DUGKS2}, the discrete velocity method (DVM) \citep{Yang2017Comparative}, the Lattice Boltzmann model (LBM) \citep{Succi-book,Shan2006Kinetic,Watari2009Velocity,Meng2011Gauss,Watari2016Is,Meng2012Lattice}, and the discrete Boltzmann method (DBM) \citep{Xu2015Progess-Combustion, Xu2016Progess-Phasesepartation, Xu2018-book, Gan2015Discrete,Lin2016Double,Lai2016Nonequilibrium,Feng2016Viscosity,Lin2017A,Lin2017Discrete,Gan2017Three}, etc. Generally, the process of simplification includes two parts.

The first part is to simplify the integral collision term on the right hand of the Boltzmann equation. Typical simplified models include the BGK model \citep{BGK1954A}, ellipsoidal statistical BGK (ES-BGK) model \citep{Holway1966New}, Shakhov model \citep{Shakhov1968Generalization}, and Rykov model \citep{Rykov1975A}, etc. Among those models the BGK is most widely used because of its clear physical significance and terse in form. However, for many cases, it has been found that quantitative results obtained from the BGK model are different from those by directly solving the original Boltzmann equation. One of the main reasons is that the corresponding hydrodynamic equation derived from the BGK by Chapman-Enskog (CE) expansion has a Prandtl number equal to unity, while it should be $2/3$ for a real monatomic gas. To improve the properties of BGK-like model, the ES-BGK and Shakhov model are developed by revising the local Maxwellian distribution function through viscous stress and heat flux, respectively. These two models retain the mathematic simplicity of the BGK model while possess an adjustable Prandtl number. Compared with the ellipsoidal distribution function in ES-BGK model, the Shakhov distribution function can be explicitly expressed by the Maxwell distribution. Consequently, it is more convenient to obtain once the Maxwell distribution function is solved.

The second part is the discretization of the particle velocity space. Some of the common approaches include discrete velocity method (DVM) \citep{UGKS2011A,Yang2016Numerical,Li2015Rarefied}, lattice Boltzmann method (LBM) \citep{Shan2006Kinetic,Meng2011Gauss,Meng2012Lattice}, and discrete Boltzmann method (DBM) \citep{Xu2015Progess-Combustion,Xu2016Progess-Phasesepartation,Xu2018-book}. Generally speaking, DVM is capable for all flow regimes but its computational efficiency is significant less than LBM, especially for higher-dimensional problems \citep{Yang2017Comparative}.
In 2009, Watari provided a three-dimensional thermal finite-difference lattice Boltzmann model to investigate the velocity slip and temperature jump phenomena \citep{Watari2009Velocity}. In that work, the key technology lies in the kinetic diffuse reflection boundary condition. Inspired by his work, Zhang et al. provided a Maxwell-type reflection boundary where the tangential momentum accommodation coefficient is introduced to describe different properties of the wall \citep{Zhang2017Velocity}. Both the diffuse and Maxwell-type reflection boundaries can accurately capture not only the velocity slip but also the Knudsen layer which can not be described by the Newton's viscosity law. In 2015, a comprehensive evaluation of higher lattice Boltzmann models is made and the applicability of higher lattice Boltzmann model to rarefied gas flows up to free molecular flow regime is verified by comparing with the analytical solutions of continuous Boltzmann equation \citep{Watari2016Is}. However, all the results are based on the BGK model. So they can not provide a proper Prandtl number, and thus can not compare with the simulations of DSMC in which hard-sphere model is usually adopted.

For a non-equilibrium flow, the Knudsen number can also be calculated by
$Kn=\tau / t^{rep}$, where $\tau$ is the relaxation time approaching local thermodynamic equilibrium, and $t^{rep}$ is a representative time scale in the flow behavior. If regard $t^{rep}$ as some time unit, then $Kn$ equals the relaxation time $\tau$. From this sense, \emph{the Knudsen number can be regarded as a kind of measure for the extent of Thermodynamic Non-Equilibrium (TNE)}.
According to the Chapman-Enskog analysis, the Navier-Stokes equations describe just the corresponding hydrodynamic model of Boltzmann equation in the continuum limit or when the system is only slightly deviated from the local thermodynamic equilibrium. The Burnett equations go a further step and contain the non-equilibrium effects in the second order of Knudsen number.
Considering the fact that general hydrodynamic equations are only for the evolutions of conserved kinetic moments of the distribution function (density, momentum and energy), while the Boltzmann equation describe the evolutions of all the conserved and non-conserved kinetic moments. In fact, not only the conserved kinetic moments but also (some of) the non-conserved kinetic moments are meaningful in describing non-equilibrium flows. The latter supplements the former in describing the specific non-equilibrium status of flow.
In recent years, DBM has been proposed to investigate both the Hydrodynamic and Thermodynamic Non-Equilibrium (HNE and TNE, respectively) behaviors in various complex flows \citep{Xu2015Progess-Combustion,Xu2016Progess-Phasesepartation,Xu2018-book}. As a coarse-grained model, its construction is based on a balance of the physical gain and computational cost.
 It has been successfully used in studying non-equilibrium phase transition \citep{Gan2015Discrete}, hydrodynamic  instabilities \citep{Lai2016Nonequilibrium,Feng2016Viscosity,Lin2017Discrete}, combustion and detonation \citep{Lin2016Double,Zhang2016Kinetic,Lin2017A}, etc. Significant new physical insights have been obtained. For example, the maximum value point of TNE strength can be used as a physical criterion to distinguish the spinnodal decomposition and domain growth stages in phase separation process \citep{Gan2015Discrete,Xu2016Progess-Phasesepartation}. The new observations on interfaces have been used to physically identify various interfaces and design corresponding interface tracking schemes \citep{Lai2016Nonequilibrium,Feng2016Viscosity}. The non-equilibrium fine structures of shock waves, have been confirmed and supplemented by other microscopic models \citep{Lin2014Polar,Kang2016Molecular,Liu2016Molecular,Liu2017Recent}.
 The high corrections between globally averaged TNE strength and density non-uniformity, between globally averaged non-organized momentum flux and velocity non-uniformity, between globally averaged non-organized energy flux and temperature non-uniformity have been used to understand the material mixing process resulted from RTI \citep{Feng2016Viscosity}.
In a recent study, the corrections between globally averaged TNE properties and non-uniformities of hydrodynamic quantities were used to understand the RMI and the competition of RMI and RTI in flow system with coexisting two kinds of instabilities \citep{Feng2018RT-RM}.
 However, all of the above works on DBM are in continuous flow regime, the application of DBM to deeper non-equilibrium or rarefied gas flows have not been well developed yet, and the quantitative comparison between the prediction of DBM and the actual situations are needed.

In this work, we develop the DBM for deeper non-equilibrium or rarefied gas flow based on Shakhov model which possesses a flexible Prandtl number. To obtain an adjustable specific heat ratio while remove the dependence of the velocity distribution function on the extra degrees of freedom, two reduced distribution functions are introduced to substitute the original velocity distribution function. A general frame work for arbitrary order discrete Shakhov-Boltzmann model is proposed based on Hermite polynomial expansion and the corresponding discrete velocity set of isotropic. A series of numerical simulations are carried out and compared with the results published before. In addition, the measurement of non-equilibrium effects is further developed and the non-equilibrium characteristics in the Kelvin-Helmholtz instability are discussed.
The remainder of this paper is organized as follows. Section $2$ demonstrates the Shakhov model and the corresponding reduced system, the nondimensionalization, the arbitrary-order discrete Shakhov-Boltzmann model, discrete scheme and boundary condition. Section $3$ verifies the new model via numerical testes including Couette flow, Fourier flow, unsteady boundary heating problem, Cavity flow, and Kelvin-Helmholtz instability. Section $4$ presents the non-equilibrium strength in $n$-th order moment space and discusses the non-equilibrium characteristics in the Kelvin-Helmholtz instability. Section $5$ concludes the present paper.

\section{Discrete Boltzmann Methods}
\subsection{Shakhov model}
To improve the performance of BGK model, in 1968, Shakhov\citep{Shakhov1968Generalization} proposed a technique to construct a model equation which provides a flexible Prandtl number based on an approximation of the Boltzmann equation for pseudo-Maxwellian molecules. The Shakhov model has a similar form with BGK except that the local equilibrium distribution function, i.e., the Maxwell distribution function, is substituted by the Shakhov distribution function $f^s$, which reads
\begin{equation}\label{Eq-Shakhov}
\frac{{\partial f}}{{\partial t}} + {v_\alpha }\frac{{\partial f}}{{\partial {r_\alpha }}} =  - \frac{1}{\tau }(f - {f^s}),
\end{equation}
with
\begin{equation}\label{Eq-fs}
{f^s} = {f^{eq}} + {f^{eq}}\left[ {(1 - Pr ){c_\alpha }{q_\alpha }\left( {\frac{{{c^2} + {\eta ^2}}}{{RT}} - 5} \right)/(5pRT)} \right],
\end{equation}
where $v_{\alpha}$ is the molecular velocity in $\alpha$ direction, $\tau$ is the reciprocal of the average collision frequency of the molecules, $c_{\alpha}$ is the thermal fluctuation of molecular velocity, ${c_\alpha } = {v_\alpha } - {u_\alpha}$ where $u_{\alpha}$ is the macroscopic velocity, $q_{\alpha}$ is the heat flux. ${c^2} = {c_\gamma}{c_\gamma}$ where Einstein summation convention is adopted. $\eta ^2$ is used to represent the energy of the extra degrees and ${\eta ^2} = {\eta _\gamma}{\eta _\gamma }$. $R$ is the ideal gas constant, $p$ and $T$ are macroscopic pressure and temperature, respectively. $f^{eq}$ indicates the Maxwell distribution function which has a form
\begin{equation}\label{Eq-feq}
{f^{eq}} = \rho {\left( {\frac{1}{{2\pi RT}}} \right)^{(D + n)/2}}\exp \left[ { - \frac{{{c^2} + {\eta ^2}}}{{2RT}}} \right],
\end{equation}
where extra degrees of freedom, i.e., molecular rotation and/or vibration, have been taken into account. $D$ indicates spacial dimension and $n$ is the number of extra degrees of freedom. In order to remove the dependence of the distribution function on the extra degrees of freedom $\eta _\gamma$, two reduced velocity distribution functions, $g$ and $h$, are often introduced and used in real computation which are defined as
\begin{equation}\label{Eq-g}
g = \int {fd{\pmb{\eta}}},
\end{equation}
\begin{equation}\label{Eq-h}
h =\frac{1}{n} \int {f\frac{{{\eta ^2}}}{2}d{\pmb{\eta }}}.
\end{equation}
Correspondingly, the two reduced local equilibrium velocity distribution functions, $g^{eq}$ and $h^{eq}$, are
\begin{equation}\label{Eq-geq}
{g^{eq}} = \int {{f^{eq}}d{\pmb{\eta }}}  = \rho {\left( {\frac{1}{{2\pi RT}}} \right)^{D/2}}\exp \left[ { - \frac{{{c^2}}}{{2RT}}} \right],
\end{equation}
and
\begin{equation}\label{Eq-heq}
{h^{eq}} = \frac{1}{n}\int {{f^{eq}}\frac{{{\eta ^2}}}{2}d{\pmb{\eta }}}  = \frac{{T}}{2}{g^{eq}}.
\end{equation}
 Similarly, the reduced Shakhov distribution functions $g^s$ and $h^s$ read
 \begin{equation}\label{Eq-gs}
{g^s} = {g^{eq}} + {g^{eq}}\left[ {(1 - Pr ){c_\alpha }{q_\alpha }\left( {\frac{{c^2}}{RT} - 4} \right)/(5pRT)} \right],
\end{equation}
 \begin{equation}\label{Eq-hs}
{h^s} = {h^{eq}} + {h^{eq}}\left[ {(1 - Pr ){c_\alpha }{q_\alpha }\left( {\frac{{c^2}}{RT} - 2} \right)/(5pRT)} \right].
\end{equation}
As a result, the evolution equation of $f$ becomes
\begin{equation}\label{Eq-Shakhov2}
\frac{\partial }{{\partial t}}\left[ {\begin{array}{*{20}{c}}
   g  \\
   h  \\
\end{array}} \right] + {v_\alpha }\frac{\partial }{{\partial {r_\alpha }}}\left[ {\begin{array}{*{20}{c}}
   g  \\
   h  \\
\end{array}} \right] =  - \frac{1}{\tau }\left[ {\begin{array}{*{20}{c}}
   {g - {g^s}}  \\
   {h - {h^s}}  \\
\end{array}} \right]
\end{equation}
The macroscopic quantities including conservative flow variables, viscous stress ($\Pi_{\alpha \beta}$), and heat flux ($q_{\alpha}$) can be expressed by $g$ and $h$

\begin{eqnarray}
  \rho &=& \int g d{\bf{v}}, \\
  \rho {u_\alpha } &=& \int g {v_\alpha }d{\bf{v}}, \\
  \rho \left( {\frac{{n + D}}{2}T + \frac{{{u^2}}}{2}} \right) &=& \int {(g\frac{{{v^2}}}{2} + n h)d{\bf{v}}}, \\
  {\Pi _{\alpha \beta }} &=& \int {(g-g^{eq}){c_\alpha }{c_\beta }d{\bf{v}}}, \\
  {q_\alpha } &=& \int ({(g-g^{eq})\frac{{{c^2}}}{2}{c_\alpha } + n (h-h^{eq}){c_\alpha })d{\bf{v}}}.
\end{eqnarray}

%
%
%
%

\subsection{Nondimensionalization}
 In this model, the reference variables are chosen as ${\rho _\infty }$, $T_\infty$, and $L_\infty$ and the following nondimensionalization are used

$\left( {{{\hat r}_\alpha },\hat L,\hat \lambda } \right) = \frac{{\left( {{{r}_\alpha }, L, \lambda } \right)}}{{{L_\infty }}}$, $\left( {{{\hat v}_\alpha },{{\hat c}_\alpha },{{\hat u}_\alpha },{{\hat \eta }_\alpha }} \right) = \frac{{\left( {{v_\alpha },{c_\alpha },{u_\alpha },{\eta _\alpha }} \right)}}{{\sqrt {R{T_\infty }} }}$, $\left( {\hat t,\hat \tau } \right) = \frac{{\left( {t,\tau } \right)}}{{{L_\infty }/\sqrt {R{T_\infty }} }}$,\\

$\hat \rho  = \frac{\rho }{{{\rho _\infty }}}$, $\hat T = \frac{T}{{{T_\infty }}}$, $\hat p = \frac{p}{{{\rho _\infty }R{T_\infty }}}$, $\hat \mu  = \frac{\mu }{{{\rho _\infty }{L_\infty }\sqrt {R{T_\infty }} }}$, $\hat \kappa  = \frac{\kappa }{{{\rho _\infty }{L_\infty }R\sqrt {R{T_\infty }} }}$,\\

$\left( {{{\hat c}_v},{{\hat c}_p}} \right) = \frac{{\left( {{c_v},{c_p}} \right)}}{R}$, $(\hat f,{\hat f^{eq}},{\hat f^s}) = \frac{{(f,{f^{eq}},{f^s})}}{{{\rho _\infty }{{(R{T_\infty })}^{ - (D + n)/2}}}}$, $(\hat g,{\hat g^{eq}},{\hat g^s}) = \frac{{(f,{f^{eq}},{f^s})}}{{{\rho _\infty }{{(R{T_\infty })}^{ - D/2}}}}$,\\

$(\hat h,{\hat h^{eq}},{\hat h^s}) = \frac{{(f,{f^{eq}},{f^s})}}{{{\rho _\infty }{{(R{T_\infty })}^{ - (D - 2)/2}}}}$,  \\
where the variables with ``$\wedge$'' sign on the left indicate the dimensionless and those without ``$\wedge$'' sign possess real physical units.

The equation of state becomes
 \begin{equation}\label{Eq-state}
\hat p = \hat \rho \hat T.
\end{equation}
The Prandtl number and Knudsen number are defined as
 \begin{equation}\label{Eq-Pr}
Pr = \frac{{{c_p}\mu }}{\kappa } = \frac{{{{\hat c}_p}\hat \mu }}{{\hat \kappa }},
\end{equation}
 \begin{equation}\label{Eq-Kn}
Kn = \frac{{\tau \sqrt {RT} }}{L} = \frac{{\hat \tau \sqrt {\hat T} }}{{\hat L}}.
\end{equation}
In the following section, unless otherwise specified, all variables are dimensionless and the ``$\wedge$'' sign will be dropped for simplicity.
\subsection{Discrete Shakhov Model}
The evolution equation of the discrete Shakhov model has a same form as Eq. (\ref{Eq-Shakhov2}), except that the velocity space is substituted by a limited number of particle velocities. As a result, the velocity distribution function $g$ ($g^s$) and $h$ ($h^s$) are replaced by the discrete distribution function $g_{ki}$ ($g_{ki}^s$) and $h_{ki}$ ($h_{ki}^s$), respectively. So the evolution equation becomes
\begin{equation}\label{Eq-discreteShakhov2}
\frac{\partial }{{\partial t}}\left[ {\begin{array}{*{20}{c}}
   {{g_{ki}}}  \\
   {{h_{ki}}}  \\
\end{array}} \right] + {v_{ki}}_\alpha \frac{\partial }{{\partial {r_\alpha }}}\left[ {\begin{array}{*{20}{c}}
   {{g_{ki}}}  \\
   {{h_{ki}}}  \\
\end{array}} \right] =  - \frac{1}{\tau }\left[ {\begin{array}{*{20}{c}}
   {{g_{ki}} - g_{ki}^s}  \\
   {{h_{ki}} - h{{_{ki}^s}}}  \\
\end{array}} \right].
\end{equation}
The discrete Shakhov distribution ($g_{ki}^{s}$) can be solved from the discrete local equilibrium distribution function ($g_{ki}^{eq}$) which is expressed as a series of Hermite polynomial up to $N$-th order\citep{Watari2016Is}
\begin{equation}\label{Eq-discrete-geq}
\begin{array}{l}
 g_{ki}^{eq} = \rho {F_k}\sum\limits_{n = 0}^N {\frac{1}{{n!}}\sum\limits_{{\eta ^n}} {\mathcal{H}_{{\eta ^n}}^{(n)}{T^{ - n/2}}} {\bf{u}}_{{\eta ^n}}^n}  \\
{\kern 16pt}  = \rho {F_k}\left[ {{\mathcal{H}^{(0)}} + \frac{1}{{1!}}{T^{ - 1/2}}\sum\limits_{{\eta _1}} {\mathcal{H}_{{\eta _1}}^{(1)}{u_{{\eta _1}}}}  + \frac{1}{{2!}}{T^{ - 1}}\sum\limits_{{\eta _1},{\eta _2}} {\mathcal{H}_{{\eta _1}{\eta _2}}^{(2)}{u_{{\eta _1}}}{u_{{\eta _2}}}} } \right. \\
 \left. {{\kern 65pt}  +  \cdots +\frac{1}{{N!}}{T^{ - N/2}}\sum\limits_{{\eta _1},{\eta _2}, \cdots {\eta _N}} {\mathcal{H}_{{\eta _1}{\eta _2} \cdots {\eta _N}}^{(N)}{u_{{\eta _1}}}{u_{{\eta _2}}} \ldots {u_{{\eta _N}}}} } \right], \\
 \end{array}
\end{equation}
where $F_k$ is the weight coefficient and $\mathcal{H}_{{\eta _1}{\eta _2} \cdots {\eta _n}}^{(n)}$ is the $n$-th order Hermite polynomial of $v_{ki\eta}$. The first several Hermite polynomials are
\begin{equation}\label{Eq-Hermite0}
  \mathcal{H}^{(0)}=1 \mathrm{,}
\end{equation}
\begin{equation}\label{Eq-Hermite1}
  \mathcal{H}^{(1)}_{\eta _1}=T^{-1/2}v_{ki\eta _1}        \mathrm{,}
\end{equation}
\begin{equation}\label{Eq-Hermite2}
  \mathcal{H}^{(2)}_{\eta _1 \eta _2}=T^{-1}v_{ki\eta _1}v_{ki\eta _2}-\delta_{\eta _1 \eta _2}  \mathrm{,}
\end{equation}
\begin{equation}\label{Eq-Hermite3}
\begin{array}{l}
    \mathcal{H}^{(3)}_{\eta _1 \eta _2 \eta _3}=T^{-3/2}v_{ki\eta _1}v_{ki\eta _2}v_{ki\eta _3}          \\
     {\kern 40pt}-T^{-1/2}(v_{ki\eta _1}\delta_{\eta _2 \eta _3}
    +v_{ki\eta _2}\delta_{\eta _1 \eta _3}
    +v_{ki\eta _3}\delta_{\eta _1 \eta _2})   \mathrm{,}
    \end{array}
\end{equation}

\begin{equation}\label{Eq-Hermite4}
\begin{array}{l}
    \mathcal{H}^{(4)}_{\eta _1 \eta _2 \eta _3 \eta _4}=T^{-2}v_{ki\eta _1}v_{ki\eta _2}v_{ki\eta _3}v_{ki\eta _4}        \\
     {\kern 50pt} -T^{-1}(
      v_{ki\eta _1}v_{ki\eta _2}\delta_{\eta _3 \eta _4}
     +v_{ki\eta _1}v_{ki\eta _3}\delta_{\eta _2 \eta _4}
     +v_{ki\eta _1}v_{ki\eta _4}\delta_{\eta _2 \eta _3}  \\ {\kern 50pt}
     +v_{ki\eta _2}v_{ki\eta _3}\delta_{\eta _1 \eta _4}
     +v_{ki\eta _2}v_{ki\eta _4}\delta_{\eta _1 \eta _3}
     +v_{ki\eta _3}v_{ki\eta _4}\delta_{\eta _1 \eta _2}
    )+{\pmb{\delta}} ^{2}_{\eta _1 \eta _2 \eta _3 \eta _4}  \mathrm{,}
    \end{array}
\end{equation}

\begin{equation}\label{Eq-Hermite5}
\begin{array}{l}
    \mathcal{H}^{(5)}_{\eta _1 \eta _2 \eta _3 \eta _4 \eta _5}=T^{-5/2}v_{ki\eta _1}v_{ki\eta _2}v_{ki\eta _3}v_{ki\eta _4}v_{ki\eta _5}      \\
    {\kern 50pt}  -T^{-3/2}(
     v_{ki\eta _1}v_{ki\eta _2}v_{ki\eta _3}\delta_{\eta _4 \eta _5}
    +v_{ki\eta _1}v_{ki\eta _2}v_{ki\eta _4}\delta_{\eta _3 \eta _5}
    +v_{ki\eta _1}v_{ki\eta _2}v_{ki\eta _5}\delta_{\eta _3 \eta _4}\\
    {\kern 50pt}
    +v_{ki\eta _1}v_{ki\eta _3}v_{ki\eta _4}\delta_{\eta _2 \eta _5}
    +v_{ki\eta _1}v_{ki\eta _3}v_{ki\eta _5}\delta_{\eta _2 \eta _4}
    +v_{ki\eta _1}v_{ki\eta _4}v_{ki\eta _5}\delta_{\eta _2 \eta _3}\\
    {\kern 50pt}
    +v_{ki\eta _2}v_{ki\eta _3}v_{ki\eta _4}\delta_{\eta _1 \eta _5}
    +v_{ki\eta _2}v_{ki\eta _3}v_{ki\eta _5}\delta_{\eta _1 \eta _4}
    +v_{ki\eta _2}v_{ki\eta _4}v_{ki\eta _5}\delta_{\eta _1 \eta _3}\\
    {\kern 50pt}
    +v_{ki\eta _3}v_{ki\eta _4}v_{ki\eta _5}\delta_{\eta _1 \eta _2})                        +T^{-1/2}(
     v_{ki\eta _1}{\pmb{\delta}} ^{2}_{\eta _2 \eta _3 \eta _4 \eta_5}
    +v_{ki\eta _2}{\pmb{\delta}} ^{2}_{\eta _1 \eta _3 \eta _4 \eta_5}\\
     {\kern 50pt}
    +v_{ki\eta _3}{\pmb{\delta}} ^{2}_{\eta _1 \eta _2 \eta _4 \eta_5}
    +v_{ki\eta _4}{\pmb{\delta}} ^{2}_{\eta _1 \eta _2 \eta _3 \eta_5}
    +v_{ki\eta _5}{\pmb{\delta}} ^{2}_{\eta _1 \eta _2 \eta _3 \eta_4}
    ) \mathrm{.}
    \end{array}
\end{equation}
where $\delta_{\eta _1 \eta _2}$ is the unit tensor and $\pmb{\delta}^{2}_{\eta _1 \eta _2 \eta _3 \eta _4} = \delta_{\eta _1 \eta _2}\delta_{\eta _3 \eta _4}+\delta_{\eta _1 \eta _3}\delta_{\eta _2 \eta _4}+\delta_{\eta _1 \eta _4}\delta_{\eta _2 \eta _3}$.

To satisfy the relations of
\begin{equation}\label{Eq-Nmodel}
\sum\limits_{k,i} {g_{ki}^{eq} {{v_{ki{\eta _1}}}{v_{ki{\eta _1}}} \cdots {v_{ki{\eta _n}}}}}  = \int {{g^{eq}}{{v_{{\eta _1}}}{v_{{\eta _1}}} \cdots {v_{{\eta _n}}}}} d{\bf{v}}{\kern 28pt} 0 \le n \le N \mathrm{,}
\end{equation}
which is defined as ``$N$-th order system'' \citep{Watari2016Is}, the discrete velocity set needs to possess the isotropic relationship up to $2N$ order. Namely, for $0\leq n \leq 2N$,
\begin{equation}\label{Eq-Moments}
  \sum\limits_{k,i} {{F_k}{T^{ - n/2}}} {v_{ki{\eta _1}}}{v_{ki{\eta _2}}} \cdots {v_{ki{\eta _n}}} = \left\{ \begin{array}{l}
 \pmb{\delta} _{{\eta _1}{\eta _2} \cdots {\eta _n}}^{n/2}, {\kern 1pt} {\kern 1pt} {\kern 1pt} {\kern 1pt} {\kern 1pt} {\kern 1pt} {\kern 1pt} {\kern 1pt} {\kern 1pt} {\kern 1pt} {\rm{if}}{\kern 1pt} {\rm{n}}{\kern 1pt} {\kern 1pt} {\kern 1pt} {\rm{is}}{\kern 1pt} {\kern 1pt} {\kern 1pt} {\kern 1pt} {\rm{even}} \\
 0, {\kern 42pt} {\rm{if}}{\kern 1pt} {\kern 1pt} {\rm{n}}{\kern 1pt} {\kern 1pt} {\kern 1pt} {\rm{is}}{\kern 1pt} {\kern 1pt} {\kern 1pt} {\kern 1pt} {\rm{odd}}. \\
 \end{array} \right.
\end{equation}

The discrete velocity model is chosen as
\begin{equation}\label{Eq-DVM1}
  {v_{ki\alpha }} = {v_k}{e_{i\alpha }} {\kern 1pt} {\kern 1pt} {\kern 1pt} {\kern 1pt} {\kern 1pt} {\kern 1pt} {\kern 1pt} {\kern 1pt} {\kern 1pt} {\kern 1pt} {\kern 1pt} (k = 0,1, \ldots ,N),
\end{equation}
which consists a rest velocity and $k$ groups moving velocities. The moving velocities of each group have the same value but different directions. In Eq. (\ref{Eq-DVM1}), $v_0=0$ corresponds to the rest velocity and ${v_{k(k \ne 0)}}$ corresponds to the value of the $k$-th group velocity. The $e_{i \alpha}$ is an unit vector indicating the direction of the discrete velocity $v_{ki\alpha}$ and can be expressed as
\begin{equation}\label{Eq-DVM2}
  ({e_{ix}},{e_{iy}}) = \left( {\cos (\frac{{i - 1}}{M}2\pi),\sin (\frac{{i - 1}}{M}2\pi)} \right) {\kern 1pt} {\kern 1pt} {\kern 1pt} {\kern 1pt} {\kern 1pt} {\kern 1pt} {\kern 1pt} {\kern 1pt} {\kern 1pt} (i = 1,2, \ldots ,M),
\end{equation}
where $M$ indicates the number of the discrete velocity in each group. This kind of unit vector set has isotropic tensors up to the ($M-1$)-th rank. Namely, for $n\leq M-1$,
\begin{equation}\label{Eq-isotropic-tensor}
  \sum\limits_i {{e_{i{\eta _1}}}{e_{i{\eta _2}}} \cdots {e_{i{\eta _n}}}}  = \left\{ \begin{array}{l}
 {A_n}\pmb{\delta} _{{\eta _1}{\eta _2} \cdots {\eta _n}}^{n/2}, {\kern 1pt} {\kern 1pt} {\kern 1pt} {\kern 1pt} {\kern 1pt} {\kern 1pt} {\kern 1pt} {\kern 1pt} {\kern 1pt} {\kern 1pt} {\kern 1pt} {\kern 1pt} {\rm{if}}{\kern 1pt} {\rm{n}}{\kern 1pt} {\kern 1pt} {\kern 1pt} {\rm{is}}{\kern 1pt} {\kern 1pt} {\kern 1pt} {\kern 1pt} {\rm{even}} \\
 0, {\kern 58pt} {\rm{if}}{\kern 1pt} {\kern 1pt} {\rm{n}}{\kern 1pt} {\kern 1pt} {\kern 1pt} {\rm{is}}{\kern 1pt} {\kern 1pt} {\kern 1pt} {\kern 1pt} {\rm{odd}} \\
 \end{array} \right.
\end{equation}
where ${A_n} = \frac{M}{{n!!}}$ and ${A_0} = M$.

Combining the requirement of Eq. (\ref{Eq-Moments}) we can get that, in order to realize the $N$-th system, (i)the number of discrete velocity $M$ in each group must be greater than $2N+1$ and (ii) the weight coefficients $F_k$ need to satisfy the following equations
\begin{equation}\label{Eq-requirementofFk}
\left\{ \begin{array}{l}
 {F_0} + \sum\limits_{k \ne 0} {{F_k}{A_0}}  = 1, \\
 \sum\limits_{k \ne 0} {{F_k}v_k^2{A_2}}  = T, \\
 \sum\limits_{k \ne 0} {{F_k}v_k^4{A_4}}  = {T^2}, \\
  \vdots  \\
 \sum\limits_{k \ne 0} {{F_k}v_k^{2N}{A_{2N}}}  = {T^N}. \\
 \end{array} \right.
\end{equation}
The $F_k$ can be solved from Eq.(\ref{Eq-requirementofFk}) and the expressions of $F_k$ for $N=$ 4, 6, 8, and 10 are presented in Ref. \citep{Watari2016Is}. The general form of $F_k$ can be expressed as
\begin{equation}\label{Eq-Fk}
  {F_k} = \frac{{\sum\limits_{n = 1}^N {{B_n}} {G_{N - n}}(\underbrace {{v_1},{v_2}, \cdots ,{v_{k - 1}},{v_{k + 1}}, \cdots ,{v_N}}_{N - 1})}}{{v_k^2\prod\limits_{n = 1,n \ne k}^N {(v_k^2 - v_n^2)} }},
\end{equation}
and
\begin{equation}\label{Eq-F0}
{F_0} = 1 - {B_0}\sum\limits_{n = 1}^N {{F_n}},
\end{equation}
where
\begin{equation}\label{Eq-Bn}
  {B_n} = \left\{ \begin{array}{l}
 M, {\kern 92pt} n = 0 \\
 {( - 1)^{n+N}}\frac{{(2n)!!}}{M}{T^n}, {\kern 28pt} n \ne 0 \\
 \end{array}, \right.
\end{equation}
and
\begin{equation}\label{Eq-GNn}
  {G_{N - n}}\left( {{x_1},{x_2}, \cdots ,{x_{N - 1}}} \right) = \left\{ \begin{array}{l}
 1, {\kern 180pt} n = N \\
 {\sum\limits_{{m_1}<{m_2}< \cdots  <{m_{N-n}}}^{N-1} {{x_{{m_1}}^2} {x_{{m_2}}^2} \cdots {x_{{m_{N-n}}}^2} } }, {\kern 31pt}   1 \le n < N \\
 \end{array}. \right.
\end{equation}
Up to now the discrete equilibrium distribution function, $g_{ki}^{eq}$, can be solved. The $h_{ki}^{eq}$ can also be obtained from $g_{ki}^{eq}$ by
\begin{equation}\label{Eq-discrete-heq}
  h_{ki}^{eq} = \frac{{T}}{2}g_{ki}^{eq} \mathrm{.}
\end{equation}
Then the corresponding discrete Shakhov distribution function, $g_{ki}^s$ and $h_{ki}^s$, can be derived by
\begin{equation}\label{Eq-gski}
  g_{ki}^s = g_{ki}^{eq} + g_{ki}^{eq}\left[ {(1 - Pr ){c_{ki\alpha}}{q_\alpha }\left( {\frac{c_{ki}^2}{T} - 4} \right)/(5pT)} \right] \mathrm{,}
\end{equation}
and
\begin{equation}\label{Eq-gski}
  h_{ki}^s = h_{ki}^{eq} + h_{ki}^{eq}\left[ {(1 - Pr ){c_{ki\alpha}}{q_\alpha }\left( {\frac{c_{ki}^2}{T} - 2} \right)/(5pT)} \right] \mathrm{,}
\end{equation}
respectively.
\subsection{Discrete schemes and boundary condition}
\begin{figure}
  \centering
  \includegraphics[width=0.8\textwidth]{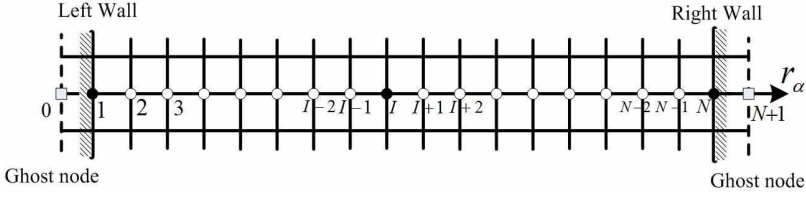}\\
  \caption{Schematic illustration of the computational grid in the $r_{\alpha}$ direction.}\label{Fig0}
\end{figure}

In order to solve Eq. (\ref{Eq-discreteShakhov2}) numerically, finite-difference method is used. The time derivatives are solved by the first-order forward scheme and the spatial derivatives are solved by the second-order upwind scheme. Then the distribution functions update can be written as
\begin{equation}\label{Eq:gkiFD}
  g_{ki}^{t + \Delta t} = g_{ki}^t - {v_{ki\alpha }}\frac{{\partial {g_{ki}}}}{{\partial {r_\alpha }}}\Delta t - \frac{1}{\tau }({g_{ki}} - g_{ki}^{s})\Delta t ,
\end{equation}
and
\begin{equation}\label{Eq:hkiFD}
  h_{ki}^{t + \Delta t} = h_{ki}^t - {v_{ki\alpha }}\frac{{\partial {h_{ki}}}}{{\partial {r_\alpha }}}\Delta t - \frac{1}{\tau }({h_{ki}} - h_{ki}^{s})\Delta t,
\end{equation}
where the spatial derivations of $g_{ki}$ and $h_{ki}$ at position $I$ (see Fig. \ref{Fig0}) are calculated by
\begin{equation}\label{Eq:2rdFDg}
\frac{{\partial {g_{ki}}}}{{\partial {r_\alpha }}} = \left\{ \begin{array}{l}
 \frac{{3{g_{ki,I}} - 4{g_{ki,I - 1}} + {g_{ki,I - 2}}}}{{2\Delta {r_\alpha }}}{\kern 14pt}  {\rm{if}}{\kern 5pt} {v_{ki\alpha }} \ge 0 \\
 \frac{{3{g_{ki,I}} - 4{g_{ki,I + 1}} + {g_{ki,I + 2}}}}{{ - 2\Delta {r_\alpha }}}{\kern 14pt}  {\rm{if}} {\kern 5pt} {v_{ki\alpha }} < 0 \\
 \end{array}, \right.
\end{equation}
and
\begin{equation}\label{Eq:2rdFDh}
\frac{{\partial {h_{ki}}}}{{\partial {r_\alpha }}} = \left\{ \begin{array}{l}
 \frac{{3{h_{ki,I}} - 4{h_{ki,I - 1}} + {h_{ki,I - 2}}}}{{2\Delta {r_\alpha }}}{\kern 14pt}  {\rm{if}}{\kern 5pt} {v_{ki\alpha }} \ge 0 \\
 \frac{{3{h_{ki,I}} - 4{h_{ki,I + 1}} + {h_{ki,I + 2}}}}{{ - 2\Delta {r_\alpha }}}{\kern 14pt}  {\rm{if}} {\kern 5pt} {v_{ki\alpha }} < 0 \\
 \end{array} , \right.
\end{equation}
respectively.

For $v_{ki \alpha}\geq 0$, Eqs. (\ref{Eq:2rdFDg}) and (\ref{Eq:hkiFD}) are applied from $I=2$ up to $I=N+1$. At the node $I=1$, the first-order backward-difference schemes
\begin{equation}\label{Eq:1stFDg}
  \frac{{\partial {g_{ki}}}}{{\partial {r_\alpha }}} = \frac{{{g_{ki,1}} - {g_{ki,0}}}}{{\Delta {r_\alpha }}} ,
\end{equation}
and
\begin{equation}\label{Eq:1stFDh}
  \frac{{\partial {h_{ki}}}}{{\partial {r_\alpha }}} = \frac{{{h_{ki,1}} - {h_{ki,0}}}}{{\Delta {r_\alpha }}},
\end{equation}
are used. At the ghost node $I=0$, the updated distribution functions are directly obtained from the diffuse reflection boundary condition instead of Eqs. (\ref{Eq:gkiFD}) and (\ref{Eq:hkiFD})
which read
\begin{equation}\label{Eq:gkiFDleft}
  g_{ki}^{t + \Delta t} = g_{ki}^{eq}(\rho_w, u_w, T_w) ,
\end{equation}
and
\begin{equation}\label{Eq:hkiFDrigth}
  h_{ki}^{t + \Delta t} = \frac{T}{2}g_{ki}^{eq}(\rho_w, u_w, T_w),
\end{equation}
where $g_{ki}^{eq}(\rho_w, u_w, T_w)$ is obtained by substituting the $\rho_w$, $u_w$, and $T_w$ into Eq. (\ref{Eq-discrete-geq}). $u_w$, and $T_w$ are velocity and temperature on the boundary, respectively. $\rho_w$ is determined so as to give a zero-mass flow normal to the boundary \citep{Watari2009Velocity}
\begin{equation}\label{Eq:rhow}
  \rho_w=-\frac{\sum\limits_{{v_{ki\alpha}} < 0}g_{ki,1}^t v_{ki\alpha}}{\sum\limits_{{v_{ki\alpha}} > 0}g_{ki}^{eq}(\rho_w=1.0,u_w,T_w)v_{ki\alpha}} ,
\end{equation}
where $g_{ki,1}^t$ is the distribution function at node $n=1$ which can be solved from Eqs. (\ref{Eq:gkiFD}) and (\ref{Eq:2rdFDg}) when $v_{ki \alpha}< 0$.

For $v_{ki \alpha}< 0$, similar treatment is adopted. At the node $I=N$, the first-order forward-difference schemes are used and the upgraded distribution functions at the ghost node $I=N+1$ are directly solved from the diffuse reflective boundary condition.

In addition, to improve the numerical stability, sometimes the high precision difference schemes such as weighted essentially nonoscillatory (WENO) \citep{Wu2013Deterministic} and non-oscillatory non-free-parameter and dissipative (NND) \citep{NND} can also be used to solve the spatial derivatives.
\section{Numerical simulations and validations}
\subsection{Validation of specific heat ratio and Prandtl number}
 From the previous introduction, we have learned that the new model possesses both adjustable specific heat ratio, $\gamma  = (D + n + 2)/(D + n)$, and Prandtl number, $Pr$. In this section, the values of $\gamma$ and $Pr$ are verified through a series of numerical examples.

 Firstly, Sod shock tube problem under two different specific heat ratios are simulated. The initial conditions are
  \begin{equation}\label{Sod-initial}
\left\{ \begin{array}{l}
{(\rho ,{u_x},{u_y},T)_L} = (1,0,0,1),\\
{(\rho ,{u_x},{u_y},T)_R} = (0.125,0,0,0.8).
\end{array} \right.
\end{equation}
Subscripts ``L'' and ``R'' indicate macroscopic variables at the left and right sides of the discontinuity. The uniform grids with $N_x \times N_y = 1000 \times 1$ are adopted. The size of the grid is $\Delta x =\Delta y = 1\times 10^{-3}$, time step is $\Delta t = 1 \times 10^{-4}$, and the relaxation time is $\tau = 5 \times 10^{-4}$. No gradient boundary conditions are used in both the left and right boundary, which mean $f_{-1}=f_0=f_1$ and $f_{N_{x+2}}=f_{N_{x+1}}=f_{N_x}$. The eighth-order Hermite expansion and 24 directions of the discrete velocity are adopted which means $N=8$ in Eq. (\ref{Eq-discrete-geq}) and $M=24$ in Eq. (\ref{Eq-DVM2}). For numerical stability, the second NND scheme is adopted to solve the space derivation in Eq. (\ref{Eq-discreteShakhov2}).

\begin{figure}
  \centering
  \includegraphics[width=0.8\textwidth]{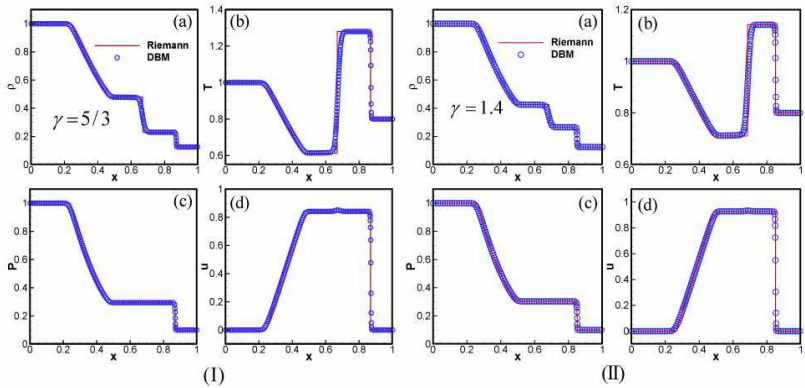}\\
  \caption{Sod shock problem with different value of specific heat ratio : (I) $\gamma =5/3$, (II) $\gamma =1.4$.}\label{Fig1}
\end{figure}

The first test has a specific heat ratio $\gamma = 5/3$ by setting $n = 1$ while the second test $\gamma = 1.4$ by setting $n=3$. Because of the different specific heat ratios, the evolution of the initial discontinuity is different over time. The macroscopic profiles at time $t=0.2$ are shown in Fig. \ref{Fig1}(I) and (II), respectively, and the Riemann analytic solutions are also plotted for comparison. It can be seen that the results of DBM are well consistent with Riemann solutions for different $\gamma$, from which we can conclude that the new model does provide an adjustable specific heat ratio.

 Secondly, the thermal Couette flows with different values of $Pr$ are simulated. Consider a gas flow between two parallel walls, one at $x=0$ and the other at $x=L$. The left wall is fixed with a temperature $T_0$ while the right wall moving with a speed $U$ in the $y$ direction with a temperature $T_1$. Based on the incompressible NS equations, the spatial distribution of temperature in the $x$ direction at steady state has an analytical solution. Specifically, the normalized temperature distribution has a relation with $Pr$ which reads \citep{Couette1,Watari2003Two}
 \begin{equation}\label{Eq-Couette}
\frac{{T - {T_0}}}{{{T_1} - {T_0}}} = \frac{x}{L} + \frac{{Pr Ec}}{2}\frac{x}{L}(1 - \frac{x}{L}),
\end{equation}
where $Ec$ is the Ecker number and is defined as $Ec = \frac{{{U^2}}}{{{C_p}({T_1} - {T_0})}}$. So this problem can be used to verify the new model with different $Pr$.

The simulations are carried out with $L=1$ and $U=0.2$, the values of temperature are $T_0=1.0$ on the left wall and $T_1=1.005$ on the right wall. The constant-pressure specific heat $C_p$ is fixed as $5/2$, various values of $Ec$ can obtained by changing the values of $T_1$. The size of the grid is $\Delta x =\Delta y = 5\times 10^{-3}$, time step is $\Delta t = 2.5 \times 10^{-4}$, and the relaxation time is $\tau = 5 \times 10^{-3}$. No slip boundary conditions are adopted in the left and right walls. The eighth-order Hermite expansion and 24 directions of the discrete velocity are adopted. For numerical stability, the second NND scheme is adopted to solve the space derivation in Eq.(\ref{Eq-discreteShakhov2}).

\begin{figure}
  \centering
  \includegraphics[width=0.8\textwidth]{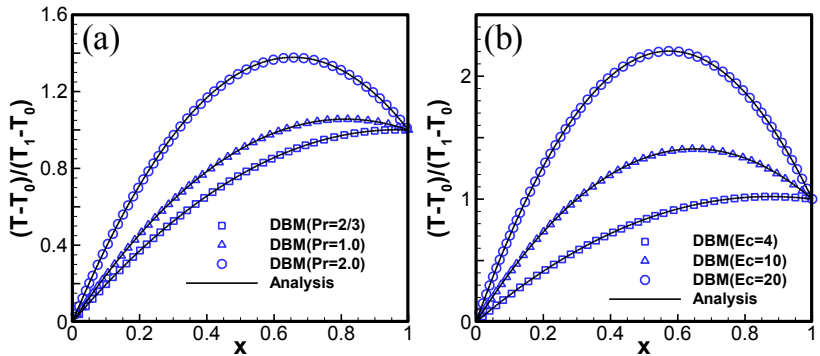}\\
  \caption{The normalized temperature distribution on steady state: (a) different values of $Pr$ with fixed $Ec=3.2$, (b) different values of $Ec$ with fixed $Pr=2/3$.}\label{Fig2}
\end{figure}

The temperature profiles at steady state are given in Fig. \ref{Fig2} where the symbols are the results of DBM and the solid lines are analytical solutions calculated from Eq. (\ref{Eq-Couette}). Figure \ref{Fig2} (a) shows the normalized temperature distribution with various values of $Pr$ when $Ec$ is fixed to $3.2$ while Figure \ref{Fig2}(b) shows the normalized temperature distribution with various values of $Ec$ when $Pr$ is fixed to $2/3$.
The results obtained from the DBM are in well agreement with analytical solutions. So it can be concluded that the new model can provide a flexible Prandtl number.

\subsection{Velocity slip and temperature jump}
Gas flows through microchannels are often encountered in MEMS and other microfluidics. Because of the larger Knudsen number, there exist significant rarefaction effects, such as the velocity slip and temperature jump near the wall. So the special boundary conditions, including the velocity-slip and temperature-jump coefficients, are needed for hydrodynamic models \citep{Manela2010Gas}. In addition, the coefficients are different for different molecular models, such as BGK and hard-sphere model \citep{Sonebook2007}. However, the discrete Boltzmann model does not need those special boundary conditions and the velocity slip and temperature jump can be naturally captured by kinetic boundary condition, i.e., the reflection of particle velocity.

In this section, the steady flows including Couette flow and Fourier flow with larger Knudsen numbers are simulated to demonstrate the velocity slip and temperature jump phenomena. The eighth-order Hermite expansion and 24 directions of the discrete velocity are adopted. The values of discrete velocities on each group are determined by the method provided in Ref. \citep{Watari2016Is}. Different Knudsen numbers are obtained by changing the values of $\tau$ according to the relationship in Eq. (\ref{Eq-Kn}).

\subsubsection{Couette flows}
Firstly, the velocity slip is demonstrated through the Couette flow test. Consider a gas flow between two parallel with the left and right wall moving with velocities $u_{wl}=-0.1$ and $u_{wr}=0.1$, respectively. The value of temperature on both walls are $T_w=1.0$.
The simulations are carried on the uniform grid with $N_x \times N_y = 200 \times 1$.
The size of the grid is $\Delta x =\Delta y = 5\times 10^{-3}$, time step is $\Delta t = 2.5 \times 10^{-4}$, and the relaxation time is determined by the Knudsen number. Diffuse reflection boundary conditions introduced in Section 2.4 are adopted. To compare with the previous results of hard-sphere gas by DSMC, the Prandtl number is set to $2/3$. In addition, the Knudsen number defined in DBM ($Kn$) has a relation with the one in DSMC ($K_D$) as follow \citep{Meng2012Lattice}
\begin{equation}\label{Eq-Kn1}
  K_D=\sqrt{\frac{\pi}{2}}Kn.
\end{equation}

\begin{figure}
  \centering
  \includegraphics[width=0.8\textwidth]{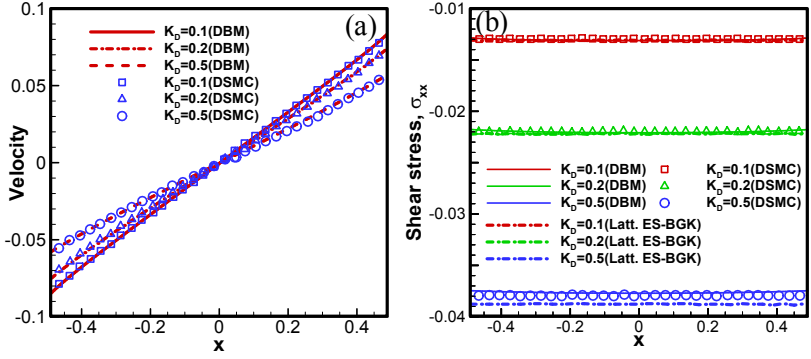}\\
  \caption{The results of Couette flow at steady state: (a) velocity profiles, the symbols are DSMC data and the lines denote DBM results; (b) viscous shear stress profiles, the symbols denote DSMC data, the dashed line represent the Lattice ES-BGK results, and the solid line represent the DBM results.}\label{Fig3}
\end{figure}

 Three kinds of Knudsen numbers are simulated and the results are compared with those of DSMC and Lattice ES-BGK \citep{Meng2012Lattice}. The velocity profiles are given in Fig. \ref{Fig3} (a). Since the results of DSMC and Lattice ES-BGK are in well agreement with each other, only the DSMC data is plotted for comparison. From Fig. \ref{Fig3} (a) we can observe that the velocity profiles are in excellent agreement with DSMC. In addition, the profiles of viscous shear stresses are shown in Fig. \ref{Fig3} (b) and the DSMC and Lattice ES-BGK results are both plotted for comparison. It can be seen that both the results of DBM and Lattice ES-BGK are well agree with DSMC for small $K_D$. However, With the increase of $K_D$, for example when $K_D=0.5$, both the results of DBM and Lattice ES-BGK have a slight deviation from DSMC. In addition, DBM and Lattice ES-BGK deviate from the DSMC in opposite direction and the results of DBM have a better consistency with DSMC than those of Lattice ES-BGK model.
\subsubsection{Fourier flows}
\begin{figure}
  \centering
  \includegraphics[width=0.8\textwidth]{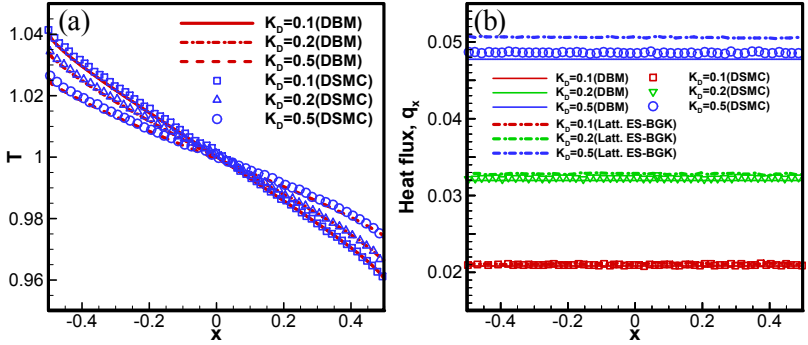}\\
  \caption{The results of Fourier flow at steady state: (a) temperature profiles, the symbols are DSMC data and the lines denote DBM results; (b) heat flux profiles, the symbols denote DSMC data, the dashed line represent the Lattice ES-BGK results, and the solid line represent the DBM results.}\label{Fig4}
\end{figure}

The temperature jump can be demonstrated through the Fourier flow test. Different from Couette flows, the left and right wall are fixed and the values of temperature are $T_{wl}=1.05$ and $T_{wr}=0.95$, respectively. The rest of the simulation conditions are same with those in the Fig. \ref{Fig3}. Three kinds of Knudsen numbers are simulated and the results are shown in Fig. \ref{Fig4}. Figure \ref{Fig4} (a) shows the temperature profiles with various $K_D$ from which we can see that the temperature jump is significant and the results have an excellence agreement with those of DSMC. The results of Lattice ES-BGK are not plotted in Fig. \ref{Fig4} (a) because they are also well agree with DSMC. The heat flux profiles are compared between DBM, Lattice ES-BGK, and DSMC in Fig. \ref{Fig4} (b). With the increase of $K_D$, both DBM and Lattice ES-BGK deviate from the DSMC data in opposite direction. However, compared with Lattice ES-BGK, the results of DBM are closer to those of DSMC.

\subsection{Unsteady boundary heating problem}
The unsteady heating processes is also a very important issue because of the common occurrence of time-varying boundary temperatures in micro-electro-mechanical and nano-electro-mechanical applications, such as microprocessor chip heating and ultrafast temperature variations in the laser industry. A series of work have been carried out by Manela and Hadjiconstantinou \citep{Manela2007On,Manela2008Gas,Manela2010Gas}. The analytical solutions for sinusoidal heating in slip flow regime were given in Ref. \citep{Manela2010Gas} and the LVDSMC method has been developed to provide efficient numerical solutions of the Boltzmann equation for all flow regimes \citep{Homolle2007A,Radtke2011Low}. In this section, we will compare our results with analytical solutions and those of LVDSMC method.

In the unsteady heating problem, two walls are fixed with a time-dependent temperature $T_w=T_0+A \mathrm{Sin}(\theta t)$ where $A$ is an amplitude which is set as $A=0.002$ in this simulations. Another non-dimensional parameter, Strouhal number ($St$), is introduced to evaluate the change rate of the temperature on the walls. $St$ is defined as $St=\theta$ and the $\theta$ has been normalized by $\sqrt{RT_{\infty}}/L_{\infty}$. When the Knudsen number is fixed, the kinetic effects can also become significant with the increase of the Strouhal number. Except the temperature on the boundary, all other parameters are set as the same with those in Fig. \ref{Fig4}.
\subsubsection{BGK model}
Firstly, the BGK gas with $Pr=1$ is simulated. There is a relation of the Knudsen number between the previous studies and this model as
\begin{equation}\label{Eq-Kn2}
  K_B=\sqrt{\frac{8}{\pi}}Kn,
\end{equation}
where $K_B$ is the definition of Knudsen number in previous studies \citep{Meng2012Lattice,Manela2010Gas} and $Kn$ is defined in Eq. (\ref{Eq-Kn}).

\begin{figure}
  \centering
  \includegraphics[width=0.8\textwidth]{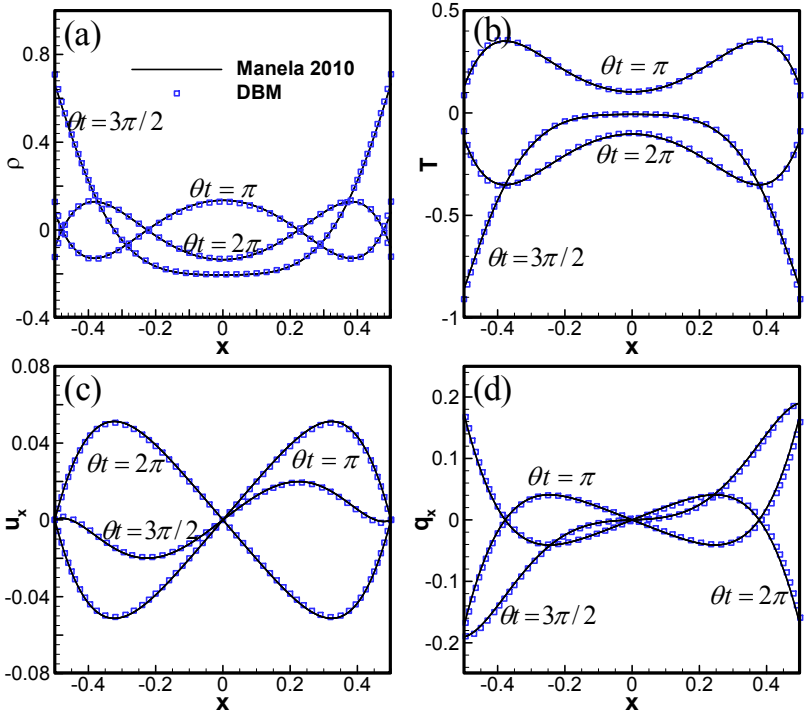}\\
  \caption{The perturbations for a BGK gas at $K_B=0.025$, $St=\pi \sqrt{2}/4$, and times $\theta t=\pi$, $\theta t=3/2\pi$, and $\theta t=2 \pi$. (a) Density, (b) temperature, (c) velocity, (d) heat flux. The solid lines indicate the analytical solutions and the symbols are DBM results.}\label{Fig5}
\end{figure}

Figure \ref{Fig5} presents the profiles of density, temperature, velocity, and heat flux perturbations at $K_B=0.025$, $St=\pi \sqrt{2}/4$, and times $\theta t=\pi$, $\theta t=3/2\pi$, and $\theta t=2 \pi$. All of the quantities have been normalized by the amplitude $A$, except that the velocity is normalized by $ASt$ \citep{Manela2010Gas}. The analytical solutions based on NS equations and first-order slip boundary condition are also plotted for comparison. The analytical expressions for density, temperature, velocity, and heat flux can be found in Eqs. (5.6), (5.1), (5.7), and (5.8) in Ref. \citep{Manela2010Gas} while the coefficient of the expression for heat flux should be corrected as $-\frac{5\widetilde{Kn}}{2Pr}$ instead of $-\frac{5 Pr\widetilde{Kn}}{2}$ in Eq. (5.8). Because the analytical solutions are well agree with LVDSMC, only the analytical solutions are compared with the DBM results. Good agreement is observed between analysis and DBM results. In addition, the temperature profiles in the Knudsen layer slightly deviate from the analytic solutions which is consistent with LVDSMC results.

\begin{figure}
  \centering
  \includegraphics[width=0.8\textwidth]{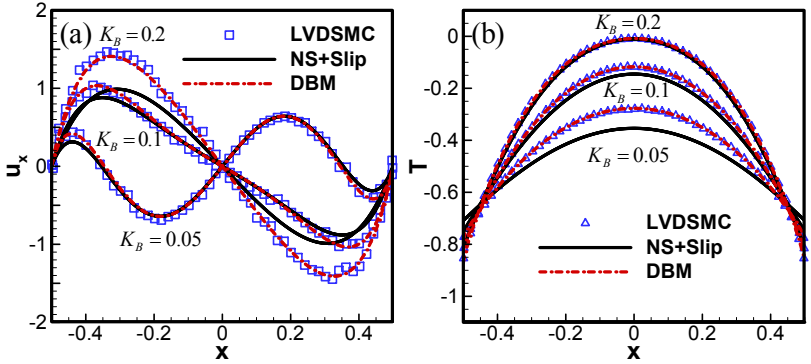}\\
  \caption{The velocity and temperature perturbations for a BGK gas at $St=\pi \sqrt{2}/4$, $\theta t=3 \pi/2$, and indicated Knudsens.(a)The velocity profiles, (b)the temperature perturbations. The symbols indicate the LVDSMC data, the solid lines repersent the analytical solutions, and the solid dot lines are the DBM results.}\label{Fig6}
\end{figure}

Figure \ref{Fig6} shows the velocity and temperature perturbations at $St=\pi \sqrt{2}/4$, $\theta t=3 \pi/2$ for various Knudsen numbers. The velocity is normalized by $A St$ and the temperature perturbations are normalized by $A$. As the $K_B$ increases into transition regime, the analytical solutions based on NS equations are not applicable anymore while the DBM can still provide accurate predictions.

Apart from the Knudsen number, the kinetic effects also depend on the Strouhal number. With increasing $St$, the wall heating possesses a higher frequency and the rarefaction effects become significant. In Ref. \citep{Meng2012Lattice}, comparisons have been made between Lattice ES-BGK and DSMC for various Strouhal and Knudsen numbers. It concluded that larger Strouhal numbers lead to discrepancies between those two models, especially for larger Knudsen numbers such as $K_B=0.5$. It has also been pointed out that this disagreement can be attributed to the moderate discrete velocity set. Based on our new model with moderate discrete velocity number($M=24$), simulations at $K_B=0.5$ for various $St$ are conducted and compared with the previous results.

\begin{figure}
  \centering
  \includegraphics[width=0.8\textwidth]{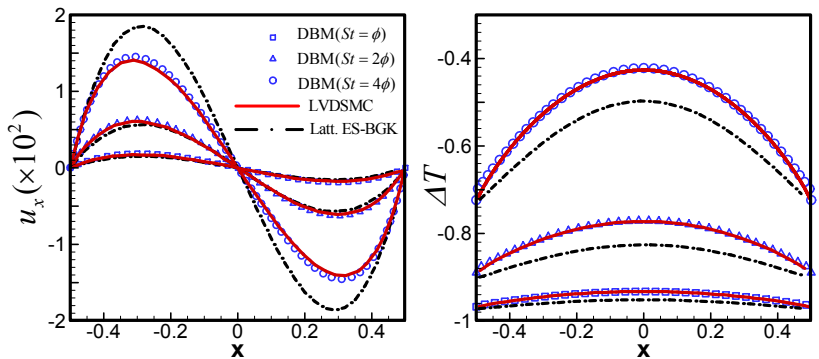}\\
  \caption{The velocity and temperature perturbations for a BGK gas at $K_B=0.5$, $\theta t=3 \pi/2$ for $St=\phi$, $St=2\phi$, and $St=4\phi$ where $\phi=\pi \sqrt{2}/16$. (a) The velocity profiles, (b) the temperature perturbations. The symbols indicate the DBM results, the solid lines repersent the LVDSMC data, and the solid dot lines are the Lattice ES-BGK results.}\label{Fig7}
\end{figure}

Figure \ref{Fig7} shows the comparison between the results of DBM, Lattice ES-BGK, and LVDSMC. Excellent agreement can be observed between the results of DBM and LVDSMC even for larger Strouhal number when Lattice ES-BGK highly deviate from the LVDSMC. In fact, the ability of the model capturing the rarefied effect for gas flow may not only depend on the number of the discrete velocity but the isotropic characteristic also needs to be taken into consideration. In addition, for compressible flow, the effects of higher order terms in the discrete equilibrium distribution function are also important. However, for this simulation, the eighth-order Hermite expansion is enough in the new model.

\subsubsection{Hard-sphere model}
The hard-sphere gas flow can also be investigated by the new model. Since the definition of Knudsen number ($K_H$) for hard-sphere model is \citep{ChingShen2005Rarefied, Bird2003}
\begin{equation}\label{Eq-Kn3}
  {K_H} = \frac{{16}}{5}\frac{\mu }{{pL}}\sqrt {\frac{{RT}}{{2\pi }}}  \mathrm{.}
\end{equation}
Combined with Eq. (\ref{Eq-Kn}), it has
\begin{equation}\label{Eq-Kn4}
  {K_H} = \frac{16}{5\sqrt{2\pi}}Kn .
\end{equation}
In addition, the Prandtl for hard-sphere gas is 2/3. The other conditions are same with those in the BGK gas.

\begin{figure}
  \centering
  \includegraphics[width=0.8\textwidth]{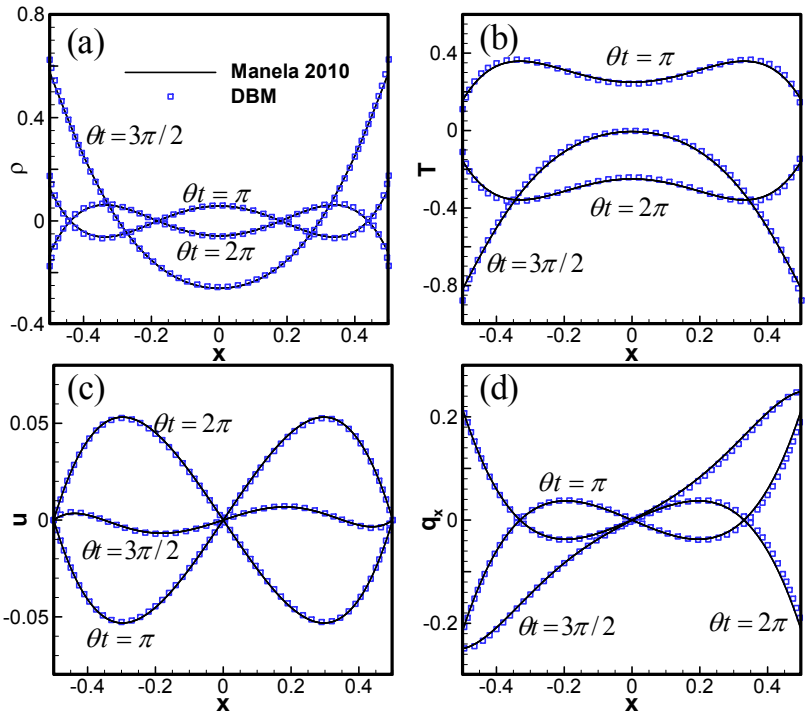}\\
  \caption{The perturbations for a hard-sphere gas at $K_H=0.025$, $St=\pi \sqrt{2}/4$, and times $\theta t=\pi$, $\theta t=3/2\pi$, and $\theta t=2 \pi$.(a)Density, (b)temperature, (c)velocity, (d)heat flux. The solid lines indicate the analytical solutions and the symbols are DBM results.}\label{Fig8}
\end{figure}

Figure \ref{Fig8} shows the profiles of density, temperature, velocity, and heat flux perturbations at $K_H=0.025$, $St=\pi \sqrt{2}/4$, and times $\theta t=\pi$, $\theta t=3/2\pi$, and $\theta t=2 \pi$. The density, temperature, and heat flux are normalized by the amplitude $A$ while the velocity is normalized by $ASt$. The DBM results are compared with the analytical solution in Ref. \citep{Manela2010Gas}. Since $K_H < 0.1$ the analytical solutions in slip flow regime are applicable. From Fig. \ref{Fig8}, it can be seen the results are well agree with the analytical solutions.

\begin{figure}
  \centering
  \includegraphics[width=0.8\textwidth]{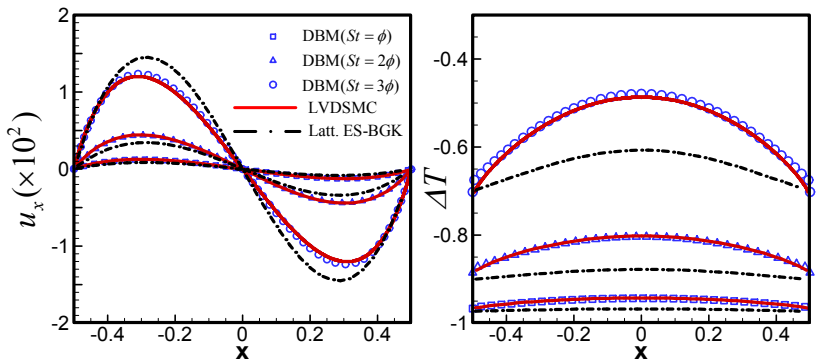}\\
  \caption{The velocity and temperature perturbations for a hard-sphere gas at $K_H=0.5$, $\theta t=3 \pi/2$ for $St=\phi$, $St=2\phi$, and $St=4\phi$ where $\phi=\pi \sqrt{2}/16$. (a)The velocity profiles, (b)the temperature perturbations. The symbols indicate the DBM results, the solid lines repersent the LVDSMC data, and the dash dot lines are the Lattice ES-BGK results.}\label{Fig9}
\end{figure}

As $K_H$ increases into transition regime, the analytical solutions are not applicable anymore. The simulations are carried out at $K_H=0.5$ for various Stroudal numbers and compared with LVDSMC data and Lattice ES-BGK results. The results at time $\theta t=3\pi /2$ are shown in Fig.\ref{Fig9}. From this figure, we can see that the results of Lattice ES-BGK largely deviate from the LVDSMC data while the results of DBM are still in excellence agreement with the LVDSMC data even for $St=4\phi$. Although it has been pointed out that increasing of the number of discrete velocities can improve the accuracy for Lattice ES-BGK, the improvement of accurate is not monotonic. The better performance of DBM is attributed to not just the number of discrete velocity but the different modeling idea. In DBM, the isotropic characteristics of discrete velocities are taken into account apart from the higher kinetic moments.

\subsection{Cavity flows}
Cavity flow problem in two dimensions is another type of boundary driven flows, which is also often encountered in non-equilibrium flows \citep{Cavity1}. Although it has been thoroughly studied and widely used as a typical benchmark for testing the model in continuum flow regime, the related researches in the slip and transition regimes are very limited. Feasible methods to investigate the non-equilibrium flow and heat transfer are DSMC \citep{Cavity-DSMC}. However, the computational cost is too high especially for those close to continuum flow regime \citep{UGKS1}. In this section, the two dimensional cavity flow in slip flow regime will be simulated using the new model and the results will be compared with those of DSMC.

A gas contained in a two-dimensional square cavity with cross section $L \times L$. The top boundary of the cavity moves with a constant horizontal velocity $U_w$ and the other three boundaries are stationary. The values of temperature in the four boundary are all fixed with $T_0$. In this simulation, the argon is chosen as the gaseous medium which is a monatomic molecule gas with atomic mass, $m=6.63 \times 10^{-26}kg$. The boundary temperature is $T_0=273K$ and the lid velocity $U_w=50m/s$. The reference temperature is chosen as $T_{\infty} = 273K$, then according to the nondimensionalization introduced in Section 2.2, the dimensionless boundary temperature is $\hat T_0 = 1.0$ and the dimensionless lid velocity is $\hat U_w =0.2097$. The reference length $L_{\infty}$ is chosen as $L$ so the dimensionless length $\hat L = 1$. The computational domain is divided uniformly into $61 \times 61$, the time step is $\Delta t = 5 \times 10^{-4}$. In the DSMC simulation, the variable hard sphere (VHS) collision model was used which has a index of viscosity $\omega = 0.81$, i.e., $\mu = \mu_{ref}(\frac{T}{T_{ref}})^{\omega}$. In order to match with the VHS model, the relaxation time in this simulation is not a constant any more but changes with temperature. According to $\mu = \tau p$ in DBM, it has $\tau = \tau_{ref}\frac{\rho_{ref}}{\rho}(\frac{T}{T_{ref}})^{\omega-1}$ where $\tau_{ref}$ is the reference relaxation time determined by Knudsen number, $\rho_{ref}$ and $T_{ref}$ are the reference density and reference temperature, respectively.  Diffuse reflection boundary conditions are adopted at four boundaries. The Prandtl number is set to $2/3$, $\gamma=5/3$, and the Knudsen number $K_H$ defined in VHS is $K_H = 0.075$ which has a relation with $Kn$ in Eq. (\ref{Eq-Kn4}).

\begin{figure}
  \centering
  \includegraphics[width=0.8\textwidth]{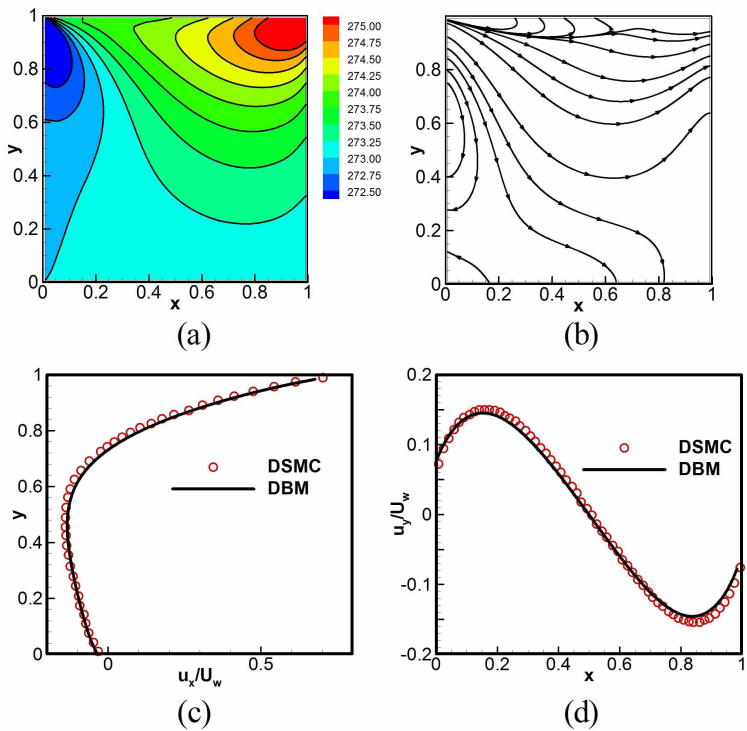}\\
  \caption{Cavity flow at $K_H=0.075$ in steady state. (a)temperature contours, (b) heat flux streamlines, (c) horizontal velocity along vertical symmetric line, (d) vertical velocity along horizontal symmetric. The symbols are DSMC data and the lines denote the DSMC results.}\label{Fig10}
\end{figure}

Figure \ref{Fig10} shows the results including plots of temperature contours, heat flux, and velocity profiles along symmetric lines. The temperature contours and the heat flux streamline plot are compared with DSMC results qualitatively. The velocity profiles normalized by $U_w$ are compared with the DSMC data quantitatively. The results of DBM are in well agreement with those of DSMC \citep{Cavity-DSMC,UGKS1}. Figure \ref{Fig10} (a) shows the temperature contour with a physical unit. It can be found that the left side is a cold region while the right side is hot. From the heat fluxes streamline plots in Fig. \ref{Fig10} (b), we can see that the direction of heat flux is mainly from the cold to the hot region which is same as the DSMC results. The heat transfer direction does not follow the gradient transport mechanism of Fourier's, which are mainly caused by the compression and thermal convection effects. Figure  \ref{Fig10} (c) and (d) show that the horizontal velocity $u_x/U_w$ along vertical symmetric line and the vertical velocity $u_y/U_w$ along horizontal symmetric are excellent agree with the DSMC data.

\subsection{Kelvin-Helmholtz instability}
The Kelvin-Helmholtz instability (KHI) occurs when two fluids have different tangential velocities and there exist a small perturbation near the interface. The KHI is ubiquity in nature and industrial process, and it is significant in both fundamental research and engineering applications such as supernova dynamics, interaction of the solar wind with the Earth's magnetosphere, aircraft engine, inertial confinement fusion, etc. \citep{KHI1,KHI2,KHI-Wang2010Combined}. In addition, KHI is also an important non-equilibrium flow and its non-equilibrium effect is particularly significant near the interface.

In this section, a two-dimensional KHI problem is simulated within a computational domain $L_x \times L_y =0.2 \times 0.2$. The initial conditions are \citep{KHI-Gan2011Lattice}
\begin{equation}\label{KHI}
\left\{ \begin{array}{l}
\rho (x) = \frac{{{\rho _L} + {\rho _R}}}{2} - \frac{{{\rho _L} - {\rho _R}}}{2}\tanh (\frac{x-0.5L_x}{{{D_\rho }}}),\\
{u_y}(x) = \frac{{{u_{yL}} + {u_{yR}}}}{2} - \frac{{{u_{yL}} - {u_{yR}}}}{2}\tanh (\frac{x-0.5L_x}{{{D_{{u_y}}}}}),\\
{P_L} = {P_R},
\end{array} \right.
\end{equation}
where $\rho_L=5.0$ and $\rho_R=2.0$ are density away from the interface of the left and right fluid, respectively, $D_{rho}=0.008$ is the widths of density transition layers, and $\tanh (x)$ indicates the hyperbolic tangent function. It has $\tanh (x)\rightarrow -1$ when $x \rightarrow -\infty$ and  $\tanh (x)\rightarrow 1$ when $x \rightarrow \infty$. Similarly, $u_{yL} = 0.5$ and $u_{yR} = -0.5$ are velocity in the $y$ direction of the left and right fluid, respectively, and $D_{u_y}=0.004$ is the widths of velocity transition layers. $P_L = P_R =2.5$ are the pressure in the left and right side. The computational domain is divided into $200 \times 200$ meshes, the size of the grid is $\Delta x =\Delta y = 1\times 10^{-3}$, time step is $\Delta t = 1 \times 10^{-5}$, and the relaxation time is $\tau = 2 \times 10^{-5}$. Besides, $\gamma =5/3$ and $Pr=1$. The velocity perturbation in the $x$ direction is set as
\begin{equation}\label{KHI2}
{u_x} = A\sin (ky)\exp \left( { - \left| {k(x - 0.5{L_x})} \right|} \right),
\end{equation}
where $A=0.02$ is the amplitude of the perturbation and $k= 2\pi/L_y$ is the wave number of the initial perturbation. Periodic boundary conditions are used in the $y$ direction and no gradient boundary condition are adopted in the $x$ direction. The fourth-order Hermite expansion and 8 directions of the discrete velocity are used to reduce the computational burden. The second NND scheme is adopted to solve the space derivation for numberical stability.
\begin{figure}
  \centering
  \includegraphics[width=0.8\textwidth]{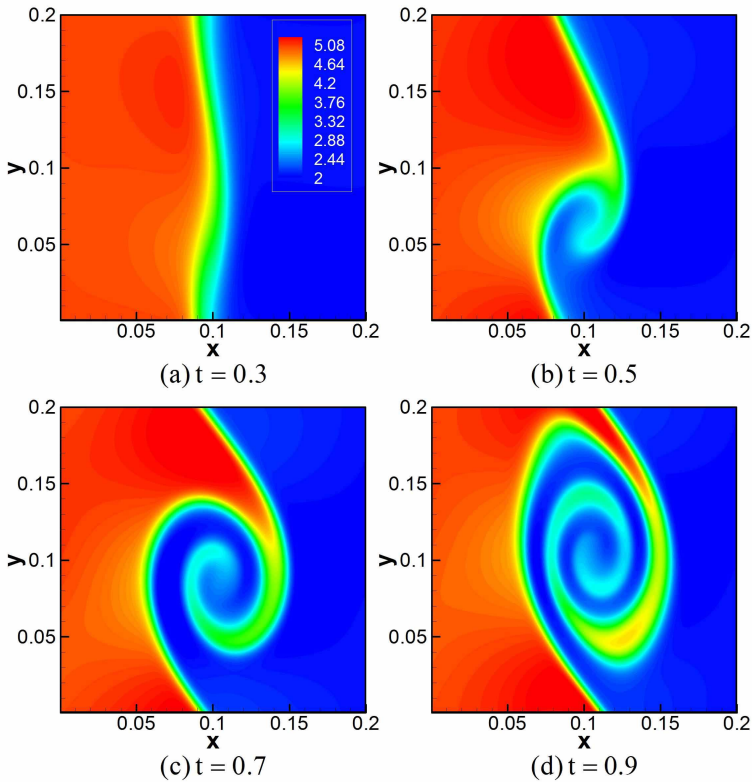}\\
  \caption{Density evolutions of KHI at different times. (a) $t=0.3$, (b) $t=0.5$, (c) $t=0.7$, (d) $t=0.9$.}\label{Fig11}
\end{figure}

Figure \ref{Fig11} shows the contours of density at four different times. It has been demonstrated that, under the condition of KHI, small perturbations along the interface undergo firstly linear then nonlinear growth stages, and finally evolve into turbulent mixing stage. Figure \ref{Fig11} (a) shows the linear growth stage when the interface is wiggling due to the initial perturbation and the velocity shear at $t=0.3$. Then the nonlinear growth stage is reached which can be found in Fig. \ref{Fig11} (b). After the initial growth stage a nicely rolled vortex is formed around the initial interface at $t=0.7$ in Fig. \ref{Fig11} (c). The vortex further rotates and grows up then a clear spiral interface can be observed at $t=0.9$ in Fig. \ref{Fig11} (d). The continuous and smooth interfaces in Fig. \ref{Fig11} indicate that the DBM can capture the interface deformation exactly.

\section{The measurements of non-equilibrium characteristic}
\subsection{Definition of the non-equilibrium strength}
In our previous work, the non-equilibrium quantities, $\pmb{\Delta}^*_n$, were presented to measure the non-equilibrium characteristics based on DBM \citep{Lin2014Polar, Gan2015Discrete,Feng2016Viscosity,Lai2016Nonequilibrium}. The definition of those non-equilibrium quantities are rewritten here
\begin{equation}\label{Detlaxingn}
\pmb{\Delta}^*_n =\mathbf{M}^*_n(f_i)-\mathbf{M}^*_n(f^{eq}_i),
\end{equation}
where $\mathbf{M}^*_n(f_i)$ and $\mathbf{M}^*_n(f^{eq}_i)$ are the $n$-th order kinetic center moment of $f_i$ and $f^{eq}_i$, respectively,
\begin{equation}\label{Mxingn}
\mathbf{M}^*_n(f_{ki}) = \sum\limits_{ki} {f_{ki} \underbrace {({{\bf{v}}_{ki}} - {\bf{u}})({{\bf{v}}_{ki}} - {\bf{u}}) \cdots ({{\bf{v}}_{ki}} - {\bf{u}})}_n}.
\end{equation}
Those non-equilibrium quantities have been proved to be helpful when we investigate the non-equilibrium flows including phase transition, Rayleigh-Taylor instability (RTI), non-equilibrium combustion and detonation, etc \citep{Gan2015Discrete, Lai2016Nonequilibrium, Feng2016Viscosity}. Some of the non-equilibrium quantities have been related to the macroscopic phenomena. For example the non-equilibrium effects can provide a physical criterion to discriminate the different stage in the phase transition \citep{Gan2015Discrete}, the $\Delta^*_{3,1,y}$ has been used to track the interface in the RTI \citep{Lai2016Nonequilibrium}, and the correlation between macroscopic non-uniformities and various global average non-equilibrium strength has been analyzed \citep{Feng2016Viscosity}.

In this work, we further define the non-equilibrium strength in the $n$-th order moment space $D^*_n$ as
\begin{equation}\label{Dn}
D^*_n = |{\pmb{\Delta}^{*}}_n|,
\end{equation}
where ${D^*_n}$ is scalar and the square of $D^*_n$ equals the sum of the squares of all the independent components in $\pmb{\Delta}^*_n$. As an example, the first several non-equilibrium strength $D^*_2$, $D^*_{3,1}$, $D^*_{3}$, and $D^*_{4,2}$ are
\begin{equation}\label{D2}
D^*_2 = \sqrt {{{(\Delta _{2,xx}^*)}^2} + {{(\Delta _{2,xy}^*)}^2} + {{(\Delta _{2,yy}^*)}^2}},
\end{equation}
\begin{equation}\label{D31}
D^*_{3,1} = \sqrt {{{(\Delta _{2,1,x}^*)}^2} + {{(\Delta _{3,1,y}^*)}^2}},
\end{equation}
\begin{equation}\label{D3}
D^*_{3} = \sqrt {{{(\Delta _{3,xxx}^*)}^2} + {{(\Delta _{3,xxy}^*)}^2}+ {{(\Delta _{3,xyy}^*)}^2}+ {{(\Delta _{3,yyy}^*)}^2}},
\end{equation}
\begin{equation}\label{D42}
D^*_{4,2} = \sqrt {{{(\Delta _{4,2,xx}^*)}^2} + {{(\Delta _{4,2,xy}^*)}^2} + {{(\Delta _{4,2,yy}^*)}^2}}.
\end{equation}
Since $\pmb{\Delta}^*_n$ is a $n$-th tensor, the independent components of $\pmb{\Delta}^*_n$ can make up a ``non-equilibrium space'' then the meaning of $D^*_n$ is the ``distance'' to the origin. Each component of $\pmb{\Delta}^*_n$ reflects the deviation from equilibrium state in a certain direction while $D^*_n$ indicates the strength of those deviation in the $n$-th order kinetic moment space. All of the independent components of $\pmb{\Delta}^*_n$, together with the $D^*_n$ make up the non-equilibrium measurement in the $n$-th order moment space. In the following section, we will show the advantage of the non-equilibrium strength $D^*_n$ to measure the non-equilibrium characteristic in the process of non-equilibrium flow, especially for those cases containing interface.
 \subsection{Non-equilibrium characterstic of the Kelvin-Helmholtz instability}
 In the evolution of KHI, the non-equilibrium effects are significant near the interface. In this section, the non-equilibrium characteristics of the KHI simulated in Fig. \ref{Fig11} will be discussed.

\begin{figure}
  \centering
  \includegraphics[width=0.8\textwidth]{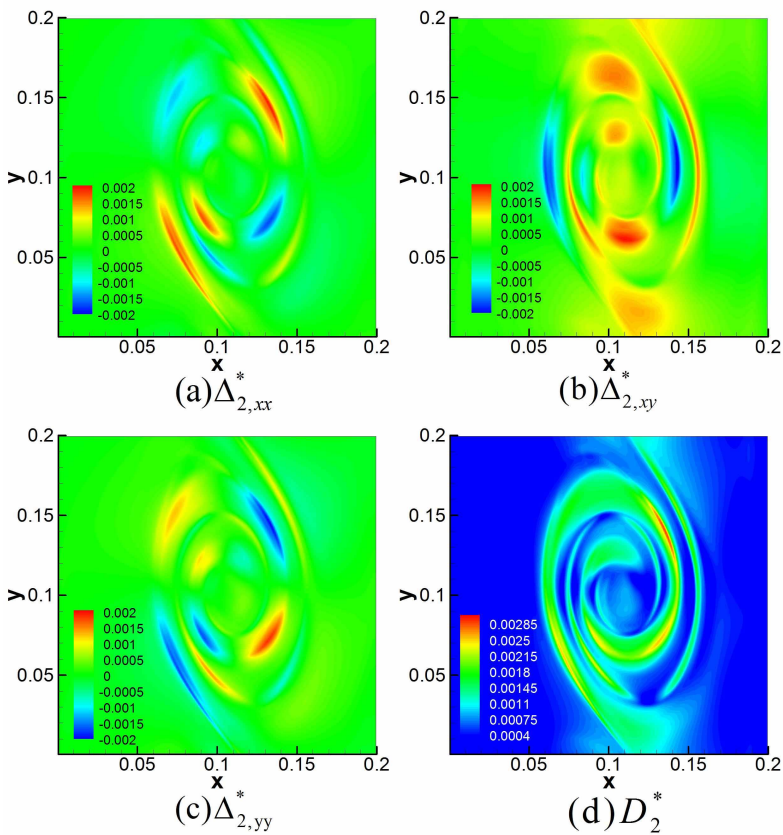}\\
  \caption{Contours of non-equilibrium characteristic in the $2$-nd order moment space at $t=0.9$. (a) $\Delta^*_{2,xx}$, (b) $\Delta^*_{2,xy}$, (c) $\Delta^*_{2,yy}$, (d) $D^*_{2}$.}\label{Fig12}
\end{figure}

 Figure \ref{Fig12} shows the contours of the independent components of $\pmb{\Delta}^{*}_2$ and the corresponding non-equilibrium strength $D^*_2$ at $t=0.9$ when a beautiful vortex is formed. Figures \ref{Fig12} (a), (b), and (c) represent the non-equilibrium characteristics near the interface in different direction which vary greatly. The value $\Delta^*_{2,xx}$ gets its maximum at which $\Delta^*_{2,yy}$ gets its minimum, in fact it has $\Delta^*_{2,xx}(x,y) + \Delta^*_{2,yy}(x,y) =0$ due to the conservation of energy. The amplitude of $\Delta^*_{2,xy}$ is close to those of $\Delta^*_{2,xx}$ and $\Delta^*_{2,yy}$, which means the shear effect is not dominant at a late stage though  the growth of the interface is mainly caused by the velocity shear at beginning. In the interface of two fluids, the value of $\Delta^*_{2,xy}$ gets its maximum (minimum) in the middle of the regions where $\Delta^*_{2,xx}$ and $\Delta^*_{2,yy}$ get their maximum (minimum). The three independent components of $\pmb{\Delta}^*_2$ constitute the completed non-equilibrium information in the $2$-nd order moment space. However, none of them are able to represent a complete interface. To get a outline of interface from the view of non-equilibrium effect, we can resort to non-equilibrium strength $D^*_{2}$. From Fig. \ref{Fig12} (d) we can see a clear spiral interface. So, the non-equilibrium strength $D^*_2$ works better for identifying the interface.

\begin{figure}
  \centering
  \includegraphics[width=0.8\textwidth]{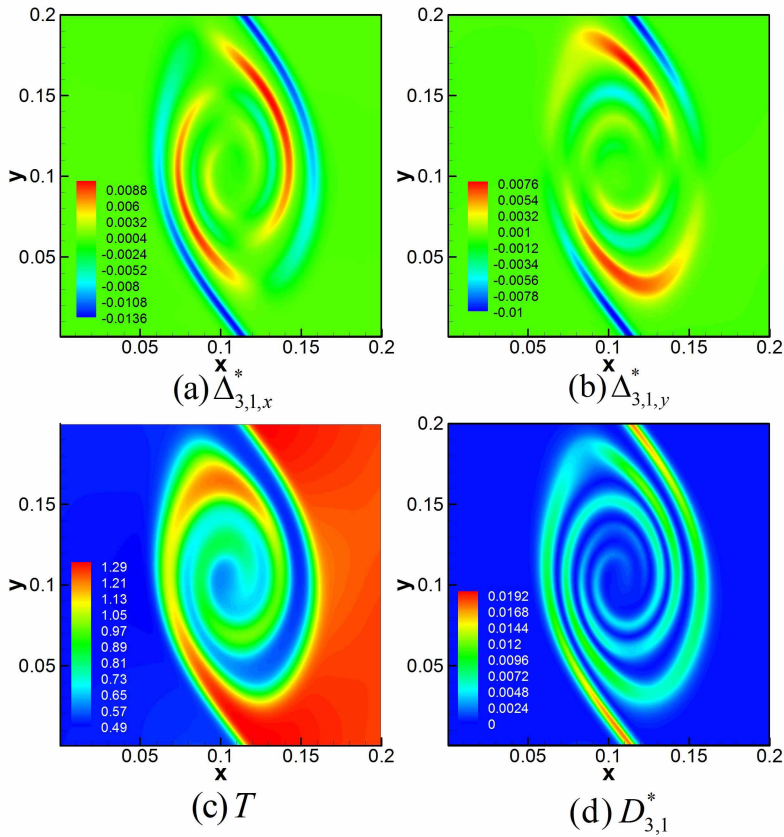}\\
  \caption{Contours of non-equilibrium effect of $\pmb{\Delta}^*_{3,1}$ at $t=0.9$. (a) $\Delta^*_{3,1,x}$, (b) $\Delta^*_{3,1,y}$, (c) $T$, (d) $D^*_{3,1}$.}\label{Fig13}
\end{figure}

Similarly, the non-equilibrium effects about the $\pmb{\Delta}^*_{3,1}$ at $t=0.9$ are presented in Fig. \ref{Fig13}. Figures \ref{Fig13} (a) and (b) show the contours of the two components $\Delta^*_{3,1,x}$ and $\Delta^*_{3,1,y}$, respectively. The contour of temperature $T$ is also plotted in Fig. \ref{Fig13} (c) for comparison. In fact, $\Delta^*_{3,1,x}$ and $\Delta^*_{3,1,y}$ correspond to the heat flux in $x$ direction and $y$ direction, respectively. From Figs. \ref{Fig13} (a) and (b), it can be found that the heat flux is negative at the outer end of the spiral, and then the positive heat flux and negative heat flux appear alternately. The closer to the center of the spiral interface, the weaker of the non-equilibrium quantities. The internal spiral interface is almost indistinguishable from $\Delta^*_{3,1,x}$ and $\Delta^*_{3,1,y}$. However, it is surprising to found $D^*_{3,1}$ provides a high resolution interface. From Fig. \ref{Fig13}, we can even see a double spiral interface which is not easy to find from the contours of density and temperature. So we conclude that the non-equilibrium strength $D^*_{3,1}$ can be well used to describe the outline of interface in the KHI simulation.

\begin{figure}
  \centering
  \includegraphics[width=0.8\textwidth]{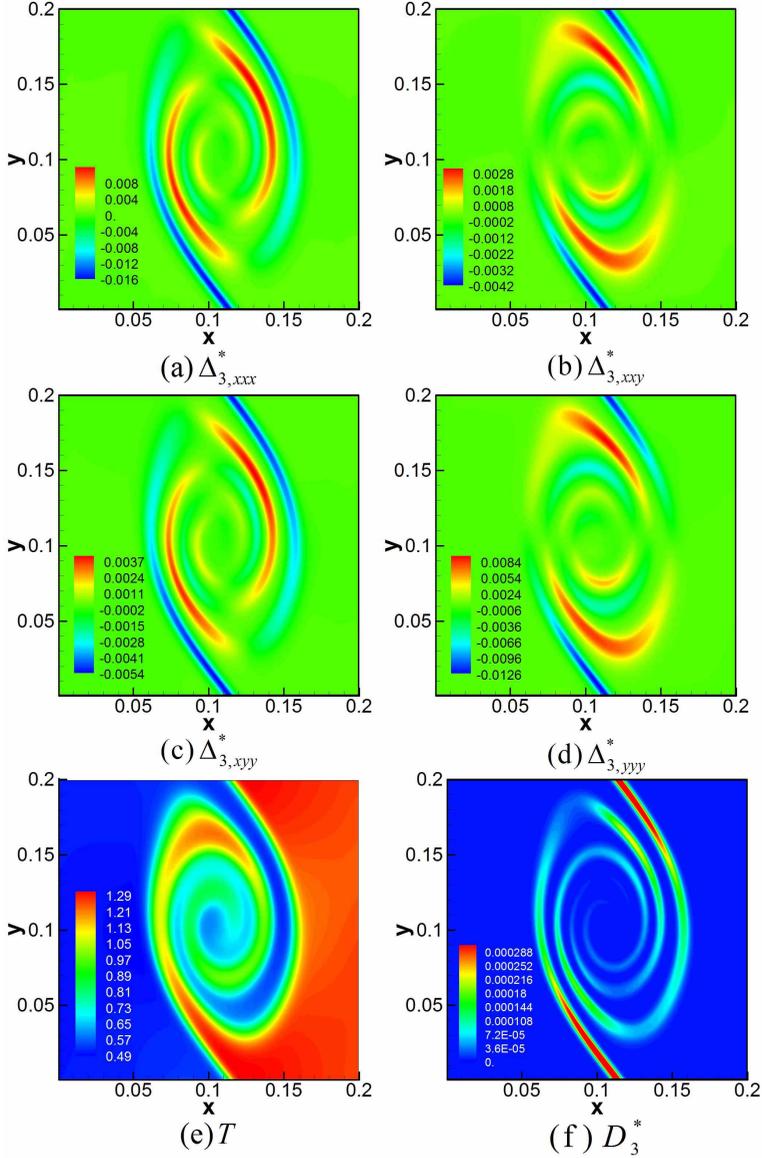}\\
  \caption{Contours of non-equilibrium characteristic in the $3$-rd order moment space at $t=0.9$. (a) $\Delta^*_{3,xxx}$, (b) $\Delta^*_{3,xxy}$, (c) $\Delta^*_{3,xyy}$, (d) $\Delta^*_{3,yyy}$, (e) $T$, (f) $D^*_{3}$.}\label{Fig14}
\end{figure}

The non-equilibrium characteristics  in the $3$-rd order moment space are also shown in Fig. \ref{Fig14}. In the two-dimensional space, $\pmb{\Delta}^*_{3}$ contains 4 independent components which are shown in Figs. \ref{Fig14} (a)-(d). The contour of temperature is given in Fig. \ref{Fig14} (e) for comparison. From the first four subgraphs, we can find the contours of $\Delta^*_{3,xxx}$ and $\Delta^*_{3,xyy}$ are very similar to each other but have different values. Generally speaking $\Delta^*_{3,xxx} > \Delta^*_{3,xyy}$. Besides, both of their contours are very similar to the contour of $\Delta^*_{3,1,x}$ in Fig. \ref{Fig13} (a), which is easy to understand because it has the relationship $\Delta^*_{3,1,x}=\frac{1}{2}(\Delta^*_{3,xxx}+\Delta^*_{3,xyy})$. Similarly, the contours of $\Delta^*_{3,xxy}$ and $\Delta^*_{3,yyy}$ are similar to each other but the value of $\Delta^*_{3,xxy}$ is less than that of $\Delta^*_{3,xyy}$. In addition, their contours are all similar to the contour of $\Delta^*_{3,1,y}$ in Fig. \ref{Fig13} (b) due to $\Delta^*_{3,1,y}=\frac{1}{2}(\Delta^*_{3,xxy}+\Delta^*_{3,yyy})$. From Fig. \ref{Fig14} (f), we can see that the non-equilibrium strength $D^*_3$ also provides a very clear double spiral interface being the same as the contour of $D^*_{3,1}$ in Fig. \ref{Fig13} (d), and the interface provided in Fig. \ref{Fig14} (f) is narrower than that in Fig. \ref{Fig13} (d), which means the $D^*_3$ can give a higher resolution interface.

\begin{figure}
  \centering
  \includegraphics[width=0.8\textwidth]{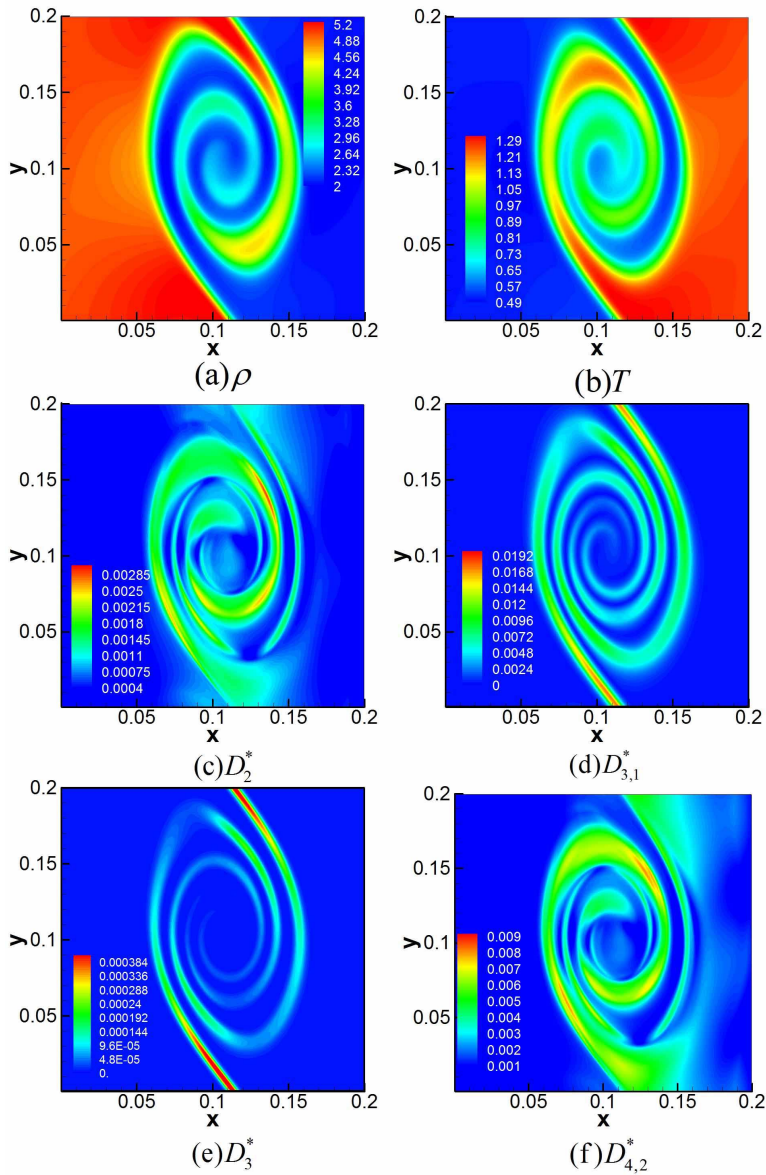}\\
  \caption{Comparison of the interfaces obtained from contours of different quantities at $t=0.9$. (a) $\rho$, (b) $T$, (c) $D^*_2$, (d) $D^*_{3,1}$, (e) $D^*_{3}$, (f) $D^*_{4,2}$.}\label{Fig15}
\end{figure}

The non-equilibrium effects about the $\pmb{\Delta}^*_{4,2}$ have a similar characteristics with those of $\pmb{\Delta}^*_{2}$. So, they are not be discussed separately. The interface obtained by macroscopic quantities and by the non-equilibrium strengthes are plotted together in Fig. \ref{Fig15}. We can find clear spiral interfaces from both the contours of density and temperature with different backgrounds in Figs. \ref{Fig15} (a) and (b). Although the interface can be well extracted from the surrounding flow field from contours of $D^*_2$ and $D^*_{4,2}$, the $D^*_3$ and $D^*_{3,1}$ can show more distinguishable interfaces. In conclusion, the non-equilibrium strength $D^*_n$ defined in this work is more appropriate to describe the interface between two fluid than the individual components of $\pmb{\Delta}^*_n$. In the KHI simulation, the non-equilibrium strength $D^*_{3}$ and $D^*_{3,1}$ are more suitable for interface representation than $D^*_{2}$ and $D^*_{4,2}$.
\section{Conclusion}
A general framework of discrete Boltzmann modeling for non-equilibrium flows is presented. The new model is based on Shakhov model, instead of BGK model. Thus, it can provide an adjustable Prandtl number. To obtain a flexible specific heat ratio while remove the dependence of velocity distribution function on the extra degrees of freedom, two reduce distribution functions are used in real computation. Arbitrary-order discrete Boltzmann model can be obtained by retaining the corresponding terms of the Hermite polynomial combining with an appropriate number of the directions of discrete velocity. The nondimensionalization process is demonstrated so that the simulation results can be recovered to cases with real physical units. A series of simulations are conducted. The results are compared with those published in previous literature. The accuracy prediction provided by the new model has been well verified for both the BGK model and hard-sphere model ranging from continuous to transition flow regime. In addition, the non-equilibrium strength $D^*_n$ is defined to further complete the measurement of non-equilibrium effect. The non-equilibrium characteristics in the Kelvin-Helmholtz instability are discussed. It has been found $D^*_n$ is more suitable to describe the overall non-equilibrium strength in the $n$-th moment space. From the contour of $D^*_n$, the interface between two fluids can be well extracted. Besides, compared with the case from $D^*_2$ and $D^*_{4,2}$,  higher resolution interfaces can be obtained from $D^*_3$ and $D^*_{3,1}$ via which clear double spiral interfaces can be identified in the simulation of Kelvin-Helmholtz instability. The observations on non-equilibrium behaviors in Kelvin-Helmholtz instability evolution are helpful for a better understanding the material mixing process resulted from hydrodynamic interfacial instability in inertial confined fusion, etc. \\

The authors are grateful to Profs. Yonghao Zhang and Jianping Meng for valuable discussions and their kind providing results of Lattice ES-BGK and LVDSMC used in the paper. This work was supported by National Natural Science Foundation of China [under Grant Nos. 11475028,11772064, 11502117, and U1530261] and Science Challenge Project (under Grant Nos. JCKY2016212A501 and TZ2016002).


\bibliographystyle{jfm}
\bibliography{jfm-instructions}

\begin{thebibliography}{62}
\expandafter\ifx\csname natexlab\endcsname\relax\def\natexlab#1{#1}\fi
\def\au#1{#1} \def\ed#1{#1} \def\yr#1{#1}\def\at#1{#1}\def\jt#1{\textit{#1}}
  \def\bt#1{#1}\def\bvol#1{\textbf{#1}} \def\vol#1{#1} \def\pg#1{#1}
  \def\publ#1{#1}\def\arxiv#1{#1}\def\org#1{#1}\def\st#1{\textit{#1}}

\bibitem[Bhatnagar {\em et~al.\/}(1954)Bhatnagar, Gross \& Krook]{BGK1954A}
{\sc \au{Bhatnagar, P.~L}, \au{Gross, E.~P} \& \au{Krook, M}} \yr{1954}  \at{A
  model for collision processes in gases. i. small amplitude processes in
  charged and neutral one-component systems}.  \jt{Physical Review}
  \bvol{94}~(3),  \pg{511--525}.

\bibitem[Bird(2003)]{Bird2003}
{\sc \au{Bird, G.}} \yr{2003} {\em Molecular Gas Dynamics and the Direct
  Simulation of Gas Flows\/}.  \publ{Oxford: Clarendon Press}.

\bibitem[Burnett(1935)]{Burnett1935The}
{\sc \au{Burnett, D.}} \yr{1935}  \at{The distribution of velocities in a
  slightly non-uniform gas}.  \jt{Proceedings of the London Mathematical
  Society}  \bvol{39}~(1),  \pg{385--430}.

\bibitem[Chen {\em et~al.\/}(2016)Chen, Xu \& Zhang]{Feng2016Viscosity}
{\sc \au{Chen, F.}, \au{Xu, A.} \& \au{Zhang, G.}} \yr{2016}  \at{Viscosity,
  heat conductivity, and prandtl number effects in the rayleigh-taylor
  instability}.  \jt{Frontiers of Physics}  \bvol{11}~(6),  \pg{183--196}.

\bibitem[Chen {\em et~al.\/}(2018)Chen, Xu \& Zhang]{Feng2018RT-RM}
{\sc \au{Chen, F.}, \au{Xu, A.} \& \au{Zhang, G.}} \yr{2018}  \at{Collaboration
  and competition between richtmyer-meshkov instability and rayleigh-taylor
  instability}.  \jt{arXiv:} .

\bibitem[Fan \& Shen(2001)]{Fan2001Statistical}
{\sc \au{Fan, J.} \& \au{Shen, C.}} \yr{2001}  \at{Statistical simulation of
  low-speed rarefied gas flows}.  \jt{Journal of Computational Physics}
  \bvol{167}~(2),  \pg{393--412}.

\bibitem[Gan {\em et~al.\/}(2018)Gan, Xu, Zhang \& Lai]{Gan2017Three}
{\sc \au{Gan, Y.}, \au{Xu, A.}, \au{Zhang, G.} \& \au{Lai, H.}} \yr{2018}
  \at{Three-dimensional discrete boltzmann models for compressible flows in and
  out of equilibrium}.  \jt{Proc IMechE Part C: J Mechanical Engineering
  Science}  \bvol{232}~(3),  \pg{477--490}.

\bibitem[Gan {\em et~al.\/}(2011)Gan, Xu, Zhang \& Li]{KHI-Gan2011Lattice}
{\sc \au{Gan, Y.}, \au{Xu, A.}, \au{Zhang, G.} \& \au{Li, Y.}} \yr{2011}
  \at{Lattice boltzmann study on kelvin-helmholtz instability: roles of
  velocity and density gradients}.  \jt{Physical Review E Statistical Nonlinear
  $\&$ Soft Matter Physics}  \bvol{83}~(2),  \pg{056704}.

\bibitem[Gan {\em et~al.\/}(2015)Gan, Xu, Zhang \& Succi]{Gan2015Discrete}
{\sc \au{Gan, Y.}, \au{Xu, A.}, \au{Zhang, G.} \& \au{Succi, S}} \yr{2015}
  \at{Discrete boltzmann modeling of multiphase flows: hydrodynamic and
  thermodynamic non-equilibrium effects}.  \jt{Soft Matter}  \bvol{11}~(26),
  \pg{5336--5345}.

\bibitem[Gard(1949)]{Grad1949}
{\sc \au{Gard, H}} \yr{1949}  \at{On the kinetic theory of rarefied gases}.
  \jt{Communications on Pure $\&$ Applied Mathematics}  \bvol{2}~(4),
  \pg{331--407}.

\bibitem[Gottscho {\em et~al.\/}(1992)Gottscho, Jurgensen \&
  Vitkavage]{Gottscho1992Microscopic}
{\sc \au{Gottscho, R.}, \au{Jurgensen, C.} \& \au{Vitkavage, D.~J}} \yr{1992}
  \at{Microscopic uniformity in plasma etching}.  \jt{Journal of Vacuum Science
  $\&$ Technology B}  \bvol{10}~(5),  \pg{2133--2147}.

\bibitem[Guo {\em et~al.\/}(2007)Guo, Shi \& Zheng]{Guo2007An}
{\sc \au{Guo, Z.}, \au{Shi, B.} \& \au{Zheng, C.}} \yr{2007}  \at{An extended
  navier-stokes formulation for gas flows in the knudsen layer near a wall}.
  \jt{Epl}  \bvol{80}~(2),  \pg{24001}.

\bibitem[Guo {\em et~al.\/}(2015)Guo, Wang \& Xu]{DUGKS2}
{\sc \au{Guo, Z.}, \au{Wang, R.} \& \au{Xu, K.}} \yr{2015}  \at{Discrete
  unified gas kinetic scheme for all knudsen number flows. ii. thermal
  compressible case}.  \jt{Physical Review E Statistical Nonlinear$\&$ Soft
  Matter Physics}  \bvol{91}~(3),  \pg{033313}.

\bibitem[Guo {\em et~al.\/}(2013)Guo, Xu \& Wang]{DUGKS1}
{\sc \au{Guo, Z.}, \au{Xu, K.} \& \au{Wang, R.}} \yr{2013}  \at{Discrete
  unified gas kinetic scheme for all knudsen number flows: low-speed isothermal
  case.}  \jt{Physical Review E Statistical Nonlinear $\&$ Soft Matter Physics}
   \bvol{88}~(3),  \pg{033305}.

\bibitem[Hasegawa {\em et~al.\/}(2004)Hasegawa, Fujimoto, Phan, RaMe, Balogh,
  Dunlop, Hashimoto \& Tandokoro]{KHI1}
{\sc \au{Hasegawa, H.}, \au{Fujimoto, M.}, \au{Phan, T.~D.}, \au{RaMe, H.},
  \au{Balogh, A.}, \au{Dunlop, M.~W.}, \au{Hashimoto, C} \& \au{Tandokoro, R}}
  \yr{2004}  \at{Transport of solar wind into earth's magnetosphere through
  rolled-up kelvin-helmholtz vortices}.  \jt{Nature}  \bvol{430}~(7001),
  \pg{755--758}.

\bibitem[Ho \& Tai(1998)]{Ho1998MICRO}
{\sc \au{Ho, C.~M.} \& \au{Tai, Y.~C.}} \yr{1998}
  \at{Micro-electro-mechanical-systems (mems) and fluid flows}.  \jt{Annual
  Review of Fluid Mechanics}  \bvol{30}~(1),  \pg{579--612}.

\bibitem[Holway~Jr(1966)]{Holway1966New}
{\sc \au{Holway~Jr, L.~H.}} \yr{1966}  \at{New statistical models for kinetic
  theory: Methods of construction}  \bvol{9}~(9),  \pg{1658--1673}.

\bibitem[Homolle \& Hadjiconstantinou(2007)]{Homolle2007A}
{\sc \au{Homolle, Thomas M.~M.} \& \au{Hadjiconstantinou, N.}} \yr{2007}  \at{A
  low-variance deviational simulation monte carlo for the boltzmann equation}.
  \jt{Journal of Computational Physics}  \bvol{226}~(2),  \pg{2341--2358}.

\bibitem[Huang {\em et~al.\/}(2012)Huang, Xu \& Yu]{UGKS1}
{\sc \au{Huang, J.}, \au{Xu, K.} \& \au{Yu, P.}} \yr{2012}  \at{A unified
  gas-kinetic scheme for continuum and rarefied flows ii: Multi-dimensional
  cases}.  \jt{Journal of Computational Physics}  \bvol{12}~(3),
  \pg{662--690}.

\bibitem[Hurricane(2009)]{KHI2}
{\sc \au{Hurricane, O.}} \yr{2009}  \at{A high energy density shock driven
  kelvin-helmholtz shear layer experiment}.  \jt{Physics of Plasmas}
  \bvol{16}~(5),  \pg{453}.

\bibitem[John {\em et~al.\/}(2011)John, Gu \& Emerson]{Cavity-DSMC}
{\sc \au{John, B.}, \au{Gu, X.} \& \au{Emerson, D.~R.}} \yr{2011}  \at{Effects
  of incomplete surface accommodation on non-equilibrium heat transfer in
  cavity flow: A parallel dsmc study}.  \jt{Computers $\&$ Fluids}
  \bvol{45}~(1),  \pg{197--201}.

\bibitem[Lai {\em et~al.\/}(2016)Lai, Xu, Zhang, Gan, Ying \&
  Succi]{Lai2016Nonequilibrium}
{\sc \au{Lai, H.}, \au{Xu, A.}, \au{Zhang, G.}, \au{Gan, Y.}, \au{Ying, Y.} \&
  \au{Succi, S.}} \yr{2016}  \at{Nonequilibrium thermohydrodynamic effects on
  the rayleigh-taylor instability in compressible flows}.  \jt{Phys.rev.e}
  \bvol{94}~(2-1).

\bibitem[Li {\em et~al.\/}(2015)Li, Peng, Zhang \& Yang]{Li2015Rarefied}
{\sc \au{Li, Z.}, \au{Peng, Ao~P.}, \au{Zhang, H.} \& \au{Yang, J.}} \yr{2015}
  \at{Rarefied gas flow simulations using high-order gas-kinetic unified
  algorithms for boltzmann model equations}.  \jt{Progress in Aerospace
  Sciences}  \bvol{74},  \pg{81--113}.

\bibitem[Lin {\em et~al.\/}(2017{\natexlab{{\em a\/}}})Lin, Luo, Fei \&
  Succi]{Lin2017A}
{\sc \au{Lin, C.}, \au{Luo, K.~Hong}, \au{Fei, L.} \& \au{Succi, S.}}
  \yr{2017{\natexlab{{\em a\/}}}}  \at{A multi-component discrete boltzmann
  model for nonequilibrium reactive flows}.  \jt{Scientific Reports}
  \bvol{7}~(1).

\bibitem[Lin {\em et~al.\/}(2016)Lin, Xu, Zhang \& Li]{Lin2016Double}
{\sc \au{Lin, C.}, \au{Xu, A.}, \au{Zhang, G.} \& \au{Li, Y.}} \yr{2016}
  \at{Double-distribution-function discrete boltzmann model for combustion}.
  \jt{Combustion $\&$ Flame}  \bvol{164},  \pg{137--151}.

\bibitem[Lin {\em et~al.\/}(2014)Lin, Xu, Zhang, Li \& Succi]{Lin2014Polar}
{\sc \au{Lin, C.}, \au{Xu, A.}, \au{Zhang, G.}, \au{Li, Y.} \& \au{Succi, S}}
  \yr{2014}  \at{Polar-coordinate lattice boltzmann modeling of compressible
  flows}.  \jt{Physical Review E Statistical Nonlinear $\&$ Soft Matter
  Physics}  \bvol{89}~(1),  \pg{013307}.

\bibitem[Lin {\em et~al.\/}(2017{\natexlab{{\em b\/}}})Lin, Xu, Zhang, Luo \&
  Li]{Lin2017Discrete}
{\sc \au{Lin, C.}, \au{Xu, A.}, \au{Zhang, G.}, \au{Luo, K.} \& \au{Li, Y.}}
  \yr{2017{\natexlab{{\em b\/}}}}  \at{Discrete boltzmann modeling of
  rayleigh-taylor instability in two-component compressible flows}.
  \jt{Physical Review E}  \bvol{96}~(5-1),  \pg{053305}.

\bibitem[Liu {\em et~al.\/}(2016{\natexlab{{\em a\/}}})Liu, Xu, Sun \&
  Cai]{UGKS2}
{\sc \au{Liu, C.}, \au{Xu, K.}, \au{Sun, Q.} \& \au{Cai, Q.}}
  \yr{2016{\natexlab{{\em a\/}}}}  \at{A unified gas-kinetic scheme for
  continuum and rarefied flows iv: Full boltzmann and model equations}.
  \jt{Journal of Computational Physics}  \bvol{314},  \pg{305--340}.

\bibitem[Liu {\em et~al.\/}(2017)Liu, Kang, Duan, Zhang \& He]{Liu2017Recent}
{\sc \au{Liu, H.}, \au{Kang, W.}, \au{Duan, H.}, \au{Zhang, P.} \& \au{He,
  X.~T.}} \yr{2017}  \at{Recent progresses on numerical investigations of
  microscopic structure of strong shock waves in fluid}.  \jt{Scientia Sinica
  Physica Mechanica $\&$ Astronomica}  \bvol{47}~(7),  \pg{070003}.

\bibitem[Liu {\em et~al.\/}(2016{\natexlab{{\em b\/}}})Liu, Kang, Zhang, Zhang,
  Duan \& He]{Kang2016Molecular}
{\sc \au{Liu, H.}, \au{Kang, W.}, \au{Zhang, Q.}, \au{Zhang, Y.}, \au{Duan, H.}
  \& \au{He, X.~T.}} \yr{2016{\natexlab{{\em b\/}}}}  \at{Molecular dynamics
  simulations of microscopicstructure of ultra strong shock waves in dense
  helium}.  \jt{Frontiers of Physics}  \bvol{11}~(6),  \pg{1--11}.

\bibitem[Liu {\em et~al.\/}(2016{\natexlab{{\em c\/}}})Liu, Zhang, Kang, Zhang,
  Duan \& He]{Liu2016Molecular}
{\sc \au{Liu, H.}, \au{Zhang, Y.}, \au{Kang, W.}, \au{Zhang, P.}, \au{Duan, H.}
  \& \au{He, X.~T.}} \yr{2016{\natexlab{{\em c\/}}}}  \at{Molecular dynamics
  simulation of strong shock waves propagating in dense deuterium with the
  effect of excited electrons}.  \jt{Physical Review E}  \bvol{95}~(2),
  \pg{023201}.

\bibitem[Manela \& Hadjiconstantinou(2007)]{Manela2007On}
{\sc \au{Manela, A.} \& \au{Hadjiconstantinou, N.~G}} \yr{2007}  \at{On the
  motion induced in a gas confined in a small-scale gap due to instantaneous
  boundary heating}.  \jt{Journal of Fluid Mechanics}  \bvol{593}~(593),
  \pg{453--462}.

\bibitem[Manela \& Hadjiconstantinou(2008)]{Manela2008Gas}
{\sc \au{Manela, Avshalom} \& \au{Hadjiconstantinou, Nicolas~G.}} \yr{2008}
  \at{Gas motion induced by unsteady boundary heating in a small-scale slab}.
  \jt{Physics of Fluids}  \bvol{20}~(11),  \pg{133}.

\bibitem[Manela \& Hadjiconstantinou(2010)]{Manela2010Gas}
{\sc \au{Manela, A.} \& \au{Hadjiconstantinou, N.~G.}} \yr{2010}  \at{Gas-flow
  animation by unsteady heating in a microchannel}.  \jt{Physics of Fluids}
  \bvol{22}~(6),  \pg{579}.

\bibitem[Meng \& Zhang(2011)]{Meng2011Gauss}
{\sc \au{Meng, J.} \& \au{Zhang, Y.}} \yr{2011}  \at{Gauss-hermite quadratures
  and accuracy of lattice boltzmann models for nonequilibrium gas flows.}
  \jt{Physical Review E Statistical Nonlinear $\&$ Soft Matter Physics}
  \bvol{83}~(2),  \pg{036704}.

\bibitem[Meng {\em et~al.\/}(2012)Meng, Zhang, Hadjiconstantinou, Radtke \&
  Shan]{Meng2012Lattice}
{\sc \au{Meng, J.}, \au{Zhang, Y.}, \au{Hadjiconstantinou, N.~G}, \au{Radtke,
  G.~A} \& \au{Shan, X.}} \yr{2012}  \at{Lattice ellipsoidal statistical bgk
  model for thermal non-equilibrium flows}.  \jt{Journal of Fluid Mechanics}
  \bvol{718}~(3),  \pg{347--370}.

\bibitem[Radtke {\em et~al.\/}(2011)Radtke, Hadjiconstantinou \&
  Wagner]{Radtke2011Low}
{\sc \au{Radtke, G.}, \au{Hadjiconstantinou, N.~G.} \& \au{Wagner, W.}}
  \yr{2011}  \at{Low-noise monte carlo simulation of the variable hard sphere
  gas}.  \jt{Physics of Fluids}  \bvol{23}~(3),  \pg{356--383}.

\bibitem[Rykov(1975)]{Rykov1975A}
{\sc \au{Rykov, V.~A.}} \yr{1975}  \at{A model kinetic equation for a gas with
  rotational degrees of freedom}.  \jt{Fluid Dynamics}  \bvol{10}~(6),
  \pg{959--966}.

\bibitem[Shakhov(1968)]{Shakhov1968Generalization}
{\sc \au{Shakhov, E.~M.}} \yr{1968}  \at{Generalization of the krook kinetic
  relaxation equation}.  \jt{Fluid Dynamics}  \bvol{3}~(5),  \pg{95--96}.

\bibitem[Shan {\em et~al.\/}(2006)Shan, Yuan \& Chen]{Shan2006Kinetic}
{\sc \au{Shan, X.}, \au{Yuan, X.} \& \au{Chen, H.}} \yr{2006}  \at{Kinetic
  theory representation of hydrodynamics: a way beyond the navier¨cstokes
  equation}.  \jt{Journal of Fluid Mechanics}  \bvol{550}~(7),  \pg{413--441}.

\bibitem[Shankar \& Deshpande(2000)]{Cavity1}
{\sc \au{Shankar, P.~N.} \& \au{Deshpande, M.~D.}} \yr{2000}  \at{Fluid
  mechanics in the driven cavity}.  \jt{Annual Review of Fluid Mechanics}
  \bvol{32}~(1),  \pg{93--136}.

\bibitem[Shen(2005)]{ChingShen2005Rarefied}
{\sc \au{Shen, C.}} \yr{2005} {\em Rarefied Gas Dynamics\/}.  \publ{Heidelberg:
  Springer}.

\bibitem[Sone(2007)]{Sonebook2007}
{\sc \au{Sone, Y.}} \yr{2007} {\em Molecular gas dynamics: theory, techniques,
  and applications\/}.  \publ{Boston: Birkhauser}.

\bibitem[Stone {\em et~al.\/}(2004)Stone, Stroock \&
  Ajdari]{Stone2004Engineering}
{\sc \au{Stone, H.~A.}, \au{Stroock, A.~D.} \& \au{Ajdari, A.}} \yr{2004}
  \at{Engineering flows in small devices: microfluidics toward a lab-on-a-chip.
  annu rev fluid mech}.  \jt{Annual Review of Fluid Mechanics}
  \bvol{36}~(:12),  \pg{381--411}.

\bibitem[Succi(2001)]{Succi-book}
{\sc \au{Succi, S.}} \yr{2001} {\em The Lattice Boltzmann Equation for Fluid
  Dynamics and Beyond\/}.  \publ{Oxford University Press, New York}.

\bibitem[Sugioka \& Cheng(2012)]{Sugioka2012Femtosecond}
{\sc \au{Sugioka, K} \& \au{Cheng, Y.}} \yr{2012}  \at{Femtosecond laser
  processing for optofluidic fabrication}.  \jt{Lab on A Chip}  \bvol{12}~(19),
   \pg{3576}.

\bibitem[Tsien(1946)]{Tsien1946Superaerodynamics}
{\sc \au{Tsien, H.}} \yr{1946}  \at{Superaerodynamics, mechanics of rarefied
  gases}.  \jt{Journal of the Aeronautical Sciences}  \bvol{13}~(12),
  \pg{653--664}.

\bibitem[Wang {\em et~al.\/}(2010)Wang, Ye \& Li]{KHI-Wang2010Combined}
{\sc \au{Wang, L.}, \au{Ye, W.} \& \au{Li, Y.}} \yr{2010}  \at{Combined effect
  of the density and velocity gradients in the combination of kelvin¨chelmholtz
  and rayleigh¨ctaylor instabilities}.  \jt{Physics of Plasmas}  \bvol{17}~(4),
   \pg{1488--12}.

\bibitem[Watari(2009)]{Watari2009Velocity}
{\sc \au{Watari, M.}} \yr{2009}  \at{Velocity slip and temperature jump
  simulations by the three-dimensional thermal finite-difference lattice
  boltzmann method}.  \jt{Physical Review E Statistical Nonlinear $\&$ Soft
  Matter Physics}  \bvol{79}~(2),  \pg{066706}.

\bibitem[Watari(2016)]{Watari2016Is}
{\sc \au{Watari, M.}} \yr{2016}  \at{Is the lattice boltzmann method applicable
  to rarefied gas flows? comprehensive evaluation of the higher-order models}.
  \jt{Journal of Fluids Engineering}  \bvol{138}~(1).

\bibitem[Watari \& Tsutahara(2003)]{Watari2003Two}
{\sc \au{Watari, M.} \& \au{Tsutahara, M.}} \yr{2003}  \at{Two-dimensional
  thermal model of the finite-difference lattice boltzmann method with high
  spatial isotropy}.  \jt{Physical Review E Statistical Nonlinear $\&$ Soft
  Matter Physics}  \bvol{67}~(2),  \pg{036306}.

\bibitem[Wu {\em et~al.\/}(2013)Wu, White, Scanlon, Reese \&
  Zhang]{Wu2013Deterministic}
{\sc \au{Wu, L.}, \au{White, C.}, \au{Scanlon, T.~J.}, \au{Reese, J.~M.} \&
  \au{Zhang, Y.}} \yr{2013}  \at{Deterministic numerical solutions of the
  boltzmann equation using the fast spectral method}.  \jt{Journal of
  Computational Physics}  \bvol{250},  \pg{27--52}.

\bibitem[Xu {\em et~al.\/}(2016)Xu, Zhang \&
  Gan]{Xu2016Progess-Phasesepartation}
{\sc \au{Xu, A.}, \au{Zhang, G.} \& \au{Gan, Y.}} \yr{2016}  \at{Progress in
  studies on discrete boltzmann modeling of phase separation process}.
  \jt{Mechanics in Engineering}  \bvol{38}~(4),  \pg{361--374}.

\bibitem[Xu {\em et~al.\/}(2015)Xu, Zhang \& Ying]{Xu2015Progess-Combustion}
{\sc \au{Xu, A.}, \au{Zhang, G.} \& \au{Ying, Y.}} \yr{2015}  \at{Progess of
  discrete boltzmann modeling and simulation of combustion system}.  \jt{Acta
  Physica Sinica}  \bvol{64}~(18),  \pg{184701}.

\bibitem[Xu {\em et~al.\/}(2018)Xu, Zhang \& Zhang]{Xu2018-book}
{\sc \au{Xu, A.}, \au{Zhang, G.} \& \au{Zhang, Y.}} \yr{2018}  \at{Discrete
  boltzmann modeling of compressible flows}.  \bt{In {\em Kinetic Theory\/}
  (ed. \ed{George~Z. Kyzas \& Athanasios~C. Mitropoulos})}, chap.~02.
  \publ{Rijeka: InTech}.

\bibitem[Xu(2001)]{Couette1}
{\sc \au{Xu, K.}} \yr{2001}  \at{A gas-kinetic bgk scheme for the navier-stokes
  equations and its connection with artificial dissipation and godunov method}.
   \jt{Journal of Computational Physics}  \bvol{171}~(1),  \pg{289--335}.

\bibitem[Xu \& Huang(2010)]{UGKS2011A}
{\sc \au{Xu, K.} \& \au{Huang, J.}} \yr{2010}  \bt{A unified gas-kinetic scheme
  for continuum and rarefied flows}. ,  \vol{vol. 229},  \pg{pp. 7747--7764}.

\bibitem[Yang {\em et~al.\/}(2016)Yang, Shu, Wu \& Wang]{Yang2016Numerical}
{\sc \au{Yang, L.~M.}, \au{Shu, C.}, \au{Wu, J.} \& \au{Wang, Y.}} \yr{2016}
  \at{Numerical simulation of flows from free molecular regime to continuum
  regime by a dvm with streaming and collision processes}.  \jt{Journal of
  Computational Physics}  \bvol{306}~(C),  \pg{291--310}.

\bibitem[Yang {\em et~al.\/}(2017)Yang, Shu, Wu \& Wang]{Yang2017Comparative}
{\sc \au{Yang, L.~M.}, \au{Shu, C.}, \au{Wu, J.} \& \au{Wang, Y.}} \yr{2017}
  \at{Comparative study of discrete velocity method and high-order lattice
  boltzmann method for simulation of rarefied flows}.  \jt{Computers $\&$
  Fluids}  \bvol{146},  \pg{125--142}.

\bibitem[Zhang(1988)]{NND}
{\sc \au{Zhang, H.}} \yr{1988}  \at{Non-oscillatory and non-free-parameter
  dissipation difference scheme}.  \jt{Acta Aerodynamica Sinica}  \bvol{6},
  \pg{143--165}.

\bibitem[Zhang {\em et~al.\/}(2018)Zhang, Xu, Zhang \& Chen]{Zhang2017Velocity}
{\sc \au{Zhang, Y.}, \au{Xu, A.}, \au{Zhang, G.} \& \au{Chen, Z.}} \yr{2018}
  \at{Discrete boltzmann method with maxwell-type boundary condition for slip
  flow}.  \jt{Commun. Theor. Phys.}  \bvol{69}~(1),  \pg{77--85}.

\bibitem[Zhang {\em et~al.\/}(2016)Zhang, Xu, Zhang, Zhu \&
  Lin]{Zhang2016Kinetic}
{\sc \au{Zhang, Y.}, \au{Xu, A.}, \au{Zhang, G.}, \au{Zhu, C.} \& \au{Lin, C.}}
  \yr{2016}  \at{Kinetic modeling of detonation and effects of negative
  temperature coefficient}.  \jt{Combustion $\&$ Flame}  \bvol{173},
  \pg{483--492}.

\end{thebibliography}

\end{document}